\documentclass[10pt]{article}

\usepackage{amsmath}
\usepackage{amssymb}
\usepackage{graphicx}
\usepackage{color}
\usepackage{subfigure}
\usepackage{rotating}
\usepackage{amsmath}
\usepackage{amssymb}
\usepackage{amsfonts}
\usepackage{float}
\usepackage{booktabs,array,dcolumn}

\usepackage{cite}

\begin{document}

%--------------------------------------------------------------------------------------------------------------------------------------------------------------------------------------------------------------------------------------------------------------------%
%--------------------------------------------------------------------------------------------------------------------------------------------------------------------------------------------------------------------------------------------------------------------%
%--------------------------------------------------------------------------------------------------------------------------------------------------------------------------------------------------------------------------------------------------------------------%

\begin{flushleft}
{\LARGE \textbf{Why {\it one-size-fits-all} vaso-modulatory interventions fail to control glioma invasion: {\it in silico} insights}}

\vspace{5mm}

J. C. L. Alfonso$^{1,*}$, A. K\"{o}hn-Luque$^{1,2,*}$, T. Stylianopoulos$^{3}$, F. Feuerhake$^{4,5}$, A. Deutsch$^{1}$ and H. Hatzikirou$^{1,6,\dag}$
\\
\vspace{3mm}

$^{(1)}$ Department for Innovative Methods of Computing, Center for Information Services and High Performance Computing, Technische Universit\"{a}t Dresden, 01062 Dresden, Germany.
\\
\vspace{1mm}

$^{(2)}$ Department of Biostatistics, Institute of Basic Medical Sciences, Faculty of Medicine, University of Oslo, 0317 Oslo, Norway.
\\
\vspace{1mm}

$^{(3)}$ Cancer Biophysics Laboratory, Department of Mechanical and Manufacturing Engineering, University of Cyprus, 1678 Nicosia, Cyprus.
\\
\vspace{1mm}

$^{(4)}$ Institute of Pathology, Medical School Hannover, 30625 Hannover, Germany.
\\
\vspace{1mm}

$^{(5)}$ Institute of Neuropathology, University Clinic Freiburg, 79117 Freiburg, Germany.
\\
\vspace{1mm}

$^{(6)}$ Department of Systems Immunology and Braunschweig Integrated Centre of Systems Biology, Helmholtz Center for Infectious Research, 38124 Braunschweig, Germany.
\\
\vspace{3mm}

$*$ These authors contributed equally to this work.
\\

$\dag$ Corresponding author: Haralampos.Hatzikrou@helmholtz-hzi.de
\\
\vspace{3mm}

The authors have declared that no competing interest exists.

\end{flushleft}

%--------------------------------------------------------------------------------------------------------------------------------------------------------------------------------------------------------------------------------------------------------------------%
%--------------------------------------------------------------------------------------------------------------------------------------------------------------------------------------------------------------------------------------------------------------------%
%--------------------------------------------------------------------------------------------------------------------------------------------------------------------------------------------------------------------------------------------------------------------%

\section*{Abstract}

There is an ongoing debate on the therapeutic potential of vaso-modulatory interventions against glioma invasion. Prominent vasculature-targeting therapies involve functional tumour-associated blood vessel deterioration and normalisation. The former aims at tumour infarction and nutrient deprivation mediated by vascular targeting agents that induce occlusion/collapse of tumour blood vessels. In contrast, the therapeutic intention of normalising the abnormal structure and function of tumour vascular networks, e.g. via alleviating stress-induced vaso-occlusion, is to improve chemo-, immuno- and radiation therapy efficacy. Although both strategies have shown therapeutic potential, it remains unclear why they often fail to control glioma invasion into the surrounding healthy brain tissue. To shed light on this issue, we propose a mathematical model of glioma invasion focusing on the interplay between the migration/proliferation dichotomy (Go-or-Grow) of glioma cells and modulations of the functional tumour vasculature. Vaso-modulatory interventions are modelled by varying the degree of vaso-occlusion. We discovered the existence of a critical cell proliferation/diffusion ratio that separates glioma invasion responses to vaso-modulatory interventions into two distinct regimes. While for tumours, belonging to one regime, vascular modulations reduce the tumour front speed and increase the infiltration width, for those in the other regime the invasion speed increases and infiltration width decreases. We show how these {\it in silico} findings can be used to guide individualised approaches of vaso-modulatory treatment strategies and thereby improve success rates.

%--------------------------------------------------------------------------------------------------------------------------------------------------------------------------------------------------------------------------------------------------------------------%
%--------------------------------------------------------------------------------------------------------------------------------------------------------------------------------------------------------------------------------------------------------------------%
%--------------------------------------------------------------------------------------------------------------------------------------------------------------------------------------------------------------------------------------------------------------------%

\vspace{5mm}
\noindent {\bf Keywords:} glioma invasion; go-or-grow mechanism; vaso-modulatory interventions; vascular occlusion and normalization; invasion speed and infiltration width; mathematical modelling.

%--------------------------------------------------------------------------------------------------------------------------------------------------------------------------------------------------------------------------------------------------------------------%
%--------------------------------------------------------------------------------------------------------------------------------------------------------------------------------------------------------------------------------------------------------------------%
%--------------------------------------------------------------------------------------------------------------------------------------------------------------------------------------------------------------------------------------------------------------------%

\section*{Introduction}

Malignant gliomas are aggressive brain tumours typically associated with a poor prognosis, sharp deterioration in the patients' quality of life and markedly low survival rates, making this disease a challenge to treat. According to the World Health Organization (WHO) \cite{Louis2007}, gliomas are classified into different categories varying from low-grade (slow-growing) to high-grade (rapidly-growing) tumours depending on their proliferative capacity and invasiveness, glioblastoma multiforme (GBM) being the most malignant form. Despite advances in surgical and medical neuro-oncology \cite{Stupp2005, Weller2010}, complete tumour resection is unlikely and subsequent recurrence is almost inevitable. A major obstacle to cure this devastating type of brain tumours is attributed to its highly invasive nature. Glioma cells have a remarkable capacity to infiltrate the surrounding normal brain tissue and migrate long distances from the tumour bed, which enables them to escape surgical resection, radiation exposure and chemotherapy \cite{Giese2003, Westphal2011, Cuddapah2014}. The persistently poor prognosis and high treatment failure rates demand more effective therapeutic strategies that should be based on a deeper mechanistic understanding of the key events triggering tumour invasion.
\vspace{2mm}

The influence of the microenvironment on the behaviour of glioma cells plays a crucial role in the resulting diffusive tumour growth and infiltration into the adjacent brain tissue. Hypoxia, the presence of abnormal and sustained low oxygen levels in the tumour tissue, strongly correlates with glioma malignancy \cite{Evans2004}. At higher glioma cell densities, tumours contain hypoxic regions with an inadequate oxygen supply due to tumour-induced vascular abnormalities. Under such oxygen-limiting conditions, glioma cells develop a wide variety of rescue mechanisms to survive and sustain proliferation. These include recruitment of new blood vessels driven by secretion of pro-angiogenic factors, modulations of cell oxygen consumption and activation of cellular migratory mechanisms to escape from poorly oxygenated regions \cite{AllalunisTurner1999, Turcotte2002, Hatzikirou2012, Hardee2012}. In particular, the ability of glioma cells to switch phenotype in response to metabolic stress may have important implications for tumour progression and resistance to therapies. For instance, the mutually exclusive switching between proliferative and migratory phenotypes experimentally observed, and known as the migration/proliferation dichotomy (or Go-or-Grow mechanism), is considered to significantly increase invasiveness in response to low oxygen levels \cite{Giese1996, Giese2003, Hatzikirou2012, Bottger2012, Bottger2015}. However, how the dynamical interplay between glioma cells and their microenvironment leads to development of hypoxic regions, as well as their global impact on glioma invasion are still not fully understood.
\vspace{2mm}

\begin{figure}[H]
\centering 
\includegraphics[width=0.95\textwidth]{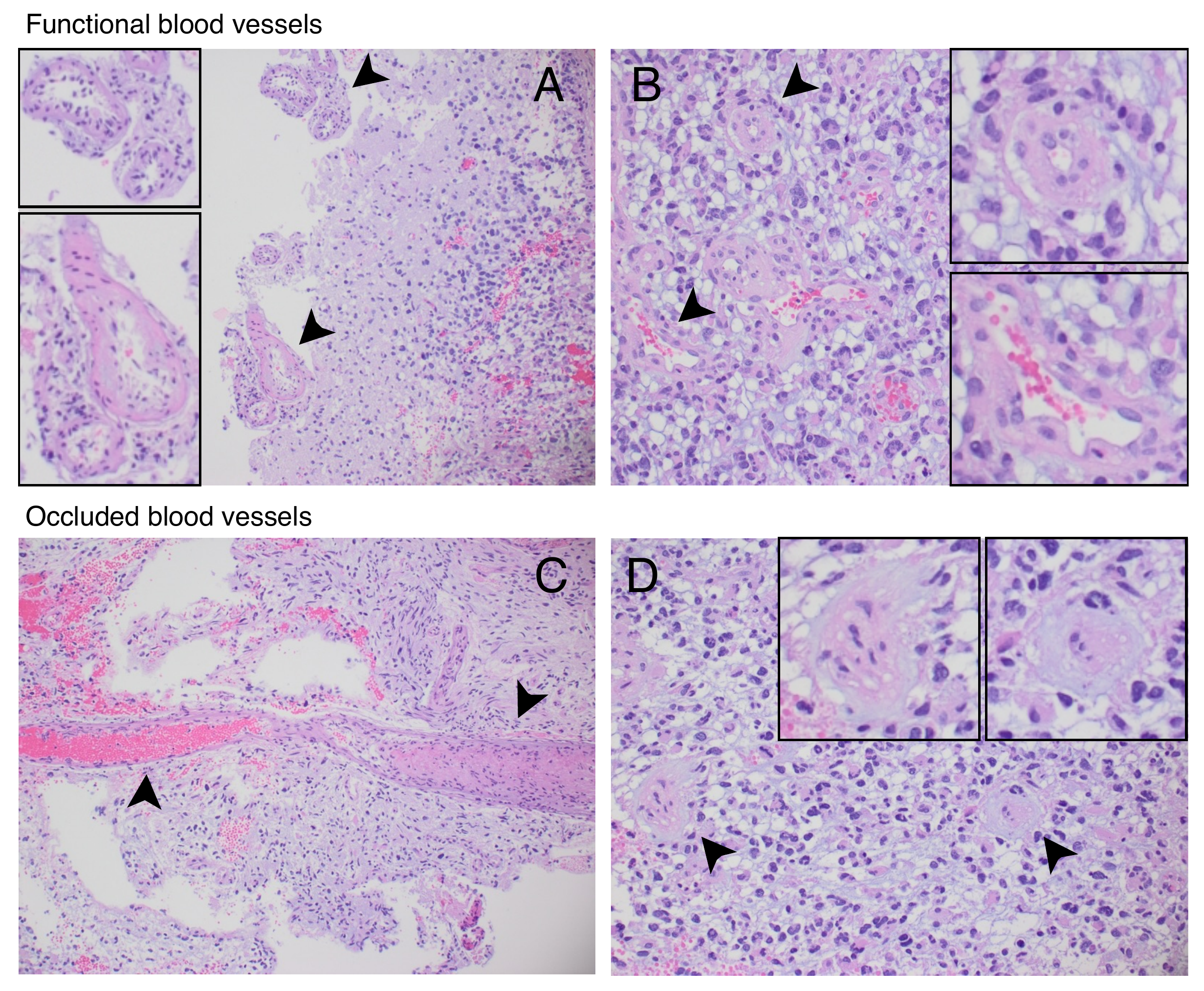}
\caption{\textbf{Histological images of functional and occluded blood vessels in malignant gliomas.} (A) From right to left brain tissue infiltrated by glioma cells with meningeal blood vessels of normal size and anatomy. (B) Atypical and not occluded intratumoural blood vessels with activated endothelium and thicker/plumper muscular layers than the normal brain vessels. (C) A longitudinal section of a large intratumoural blood vessel with a not obliterated part filled with blood (left) and an occluded part (right). (D) Thrombotic occlusion in small intratumoural blood vessels. The arrowheads point to blood vessels which are magnified in the corresponding subfigures.}
\label{fig1}
\end{figure}

A particularly important component of the tumour microenvironment is the vasculature. There exist various positive and negative feedback mechanisms between glioma cells and the vasculature. Gliomas are reported as highly vascularised neoplasias \cite{Jain2007, Swanson2011}, where excessive vascularisation is induced by a wide range of pro-angiogenic factors \cite{Carmeliet2011, Weis2011}. However, over-expression of pro-angiogenic factors produced by hypoxic glioma cells is commonly observed and results in local vascular hyperplasia with defective blood vessels. Such morphological abnormalities in the vasculature are a common feature of gliomas, where blood vessels have significantly larger diameters and thicker basement membranes than those in normal brain tissue \cite{Jain2007}, see Figure~\ref{fig1}(A,B). Moreover, vaso-occlusive events have been reported to initiate a hypoxia/necrosis cycle influencing the dynamical balance between migration and proliferation of glioma cells. In fact, different pathological and experimental observations suggest that vaso-occlusion could readily explain the rapid peripheral expansion and diffusely infiltrative growth behaviour of malignant gliomas \cite{Brat2004b, Rong2009}. Blood vessel occlusion can mainly occur due to increased mechanical pressure exerted on them by tumour cells or induced by intravascular pro-thrombotic mechanisms \cite{Brat2004a, Stamper2010}, see Figure~\ref{fig1}(C,D). Occluded or collapsed blood vessels induce perivascular tumour hypoxia and favour glioma cell migration towards better oxygenated regions. This fact has been linked to waves of hypoxic glioma cells actively migrating away from oxygen-deficient regions leading to pseudopalisade formation \cite{Brat2004a, Brat2004b, Rong2006, Rong2009}. Since hypoxia-induced migration is recognised to support further neoplastic dissemination, investigating the overall effect of vaso-modulatory interventions on the tumour front speed and infiltration width turns crucial.

The high degree of angiogenesis and vascular pathologies observed in malignant gliomas have been the target of several vaso-modulatory strategies \cite{Jain2014a, Wick2015}. Current clinical and preclinical findings suggest that angiogenesis inhibitors alone, with the potential to starve glioma cells, have limited efficacy in terms of tumour shrinkage, functional vasculature destruction and patient survival \cite{Ebos2011, Jayson2012, Jain2013}. Furthermore, anti-angiogenic factors as inhibitors of neovascularisation are also restricted by transient effects and development of therapy resistance \cite{Duda2007}. Instead, improved tumour vascularisation, either via normalisation or due to a stress alleviation strategy based on reopening compressed blood vessels, is an emerging concept expected to reduce tumour hypoxia, improve perfusion and enhance the delivery of cytotoxic drugs and radiotherapy efficacy \cite{Jain2001, Jain2005, Stylianopoulos2013, Jain2014a}. Recent evidences indicate that judicious application of an anti-angiogenic therapy may normalise the structure and function of tumour vasculature \cite{Jain2001, Jain2005, Jain2013}, where potential benefits are schedule- and patient-dependent \cite{Sorensen2012, Batchelor2013}. Although vasculature-targeting interventions could provide therapeutic benefits, further mechanistic insights into glioma invasion responses are still needed to improve treatment outcomes and patient survival \cite{Stylianopoulos2013, Jain2014a}.
\vspace{2mm}

In this work, we propose a mathematical model of reaction-diffusion type that is based on well-supported biological assumptions for the growth of vascularised gliomas. In particular, we focus on the interplay between the migration/proliferation dichotomy of glioma cells and modulations of functional tumour vasculature. Mathematical modelling has the potential to improve our understanding of the complex biology of tumours and their interactions with the microenvironment, as well as may help to design more effective and personalised therapeutic strategies \cite{Anderson2008, Byrne2010, Chauviere2012, Martinez2012, Baldock2013, Alfonso2014a, Alfonso2014, Hatzikirou2015, Reppas2015}. Several mathematical models have been developed to identify mechanisms that facilitate proliferation and migration of glioma cells \cite{Tracqui1995, Woodward1996, Burgess1997, Swanson2000, Swanson2002a, Swanson2002b, Swanson2003, Frieboes2007, Swanson2008, Swanson2011, Martinez2012, Gerlee2012}, see also \cite{Hatzikirou2005, Harpold2007} for reviews. Most of these models have been formulated to study glioma invasion based exclusively on cell diffusion and proliferation rates \cite{Tracqui1995, Woodward1996, Burgess1997, Swanson2000, Swanson2002b}. Among modelling results, interpretation of glioma growth patterns compared to clinical data \cite{Swanson2000, Swanson2002b}, as well as plausible predictions of the success or failure of different treatment techniques have been reported \cite{Tracqui1995, Woodward1996, Swanson2002a, Swanson2003, Swanson2008, Rockne2010}. Recently, different models including the influence of tumour microenvironmental conditions such as hypoxia, necrosis and angiogenesis have been developed \cite{Swanson2011, Martinez2012, Gerlee2012}. However, the role of vaso-occlusion in glioma invasion, considering the Go-or-Grow mechanism, has not been addressed so far. Accordingly, we intend to generate insights into the effects of vaso-modulatory interventions on tumour front speed and infiltration width. The main aim is to use the better understanding to investigate the potential of personalised therapeutic protocols. To that end, we begin by defining the biological assumptions taken into account when developing our glioma-vasculature interplay model. We then investigate the effect of modulations of cell oxygen consumption and vaso-occlusion rates in glioma invasion. We show that one-size-fits-all vaso-modulatory interventions should be expected to fail to control glioma growth and lead to a trade-off between tumour front speed and infiltration width. The model results provide a better understanding of glioma-microenvironment interactions, and it is therefore suited for analysing the potential success or failure of vaso-modulatory treatment strategies. We conclude with a discussion of the main implications of our model results in designing novel personalised therapeutic protocols.

%--------------------------------------------------------------------------------------------------------------------------------------------------------------------------------------------------------------------------------------------------------------------%
%--------------------------------------------------------------------------------------------------------------------------------------------------------------------------------------------------------------------------------------------------------------------%
%--------------------------------------------------------------------------------------------------------------------------------------------------------------------------------------------------------------------------------------------------------------------%

\section*{Materials and Methods}

%--------------------------------------------------------------------------------------------------------------------------------------------------------------------------------------------------------------------------------------------------------------------%

\subsection*{A glioma-vasculature interplay model}

\begin{figure}[H]
\centering 
\includegraphics[width=0.8\textwidth]{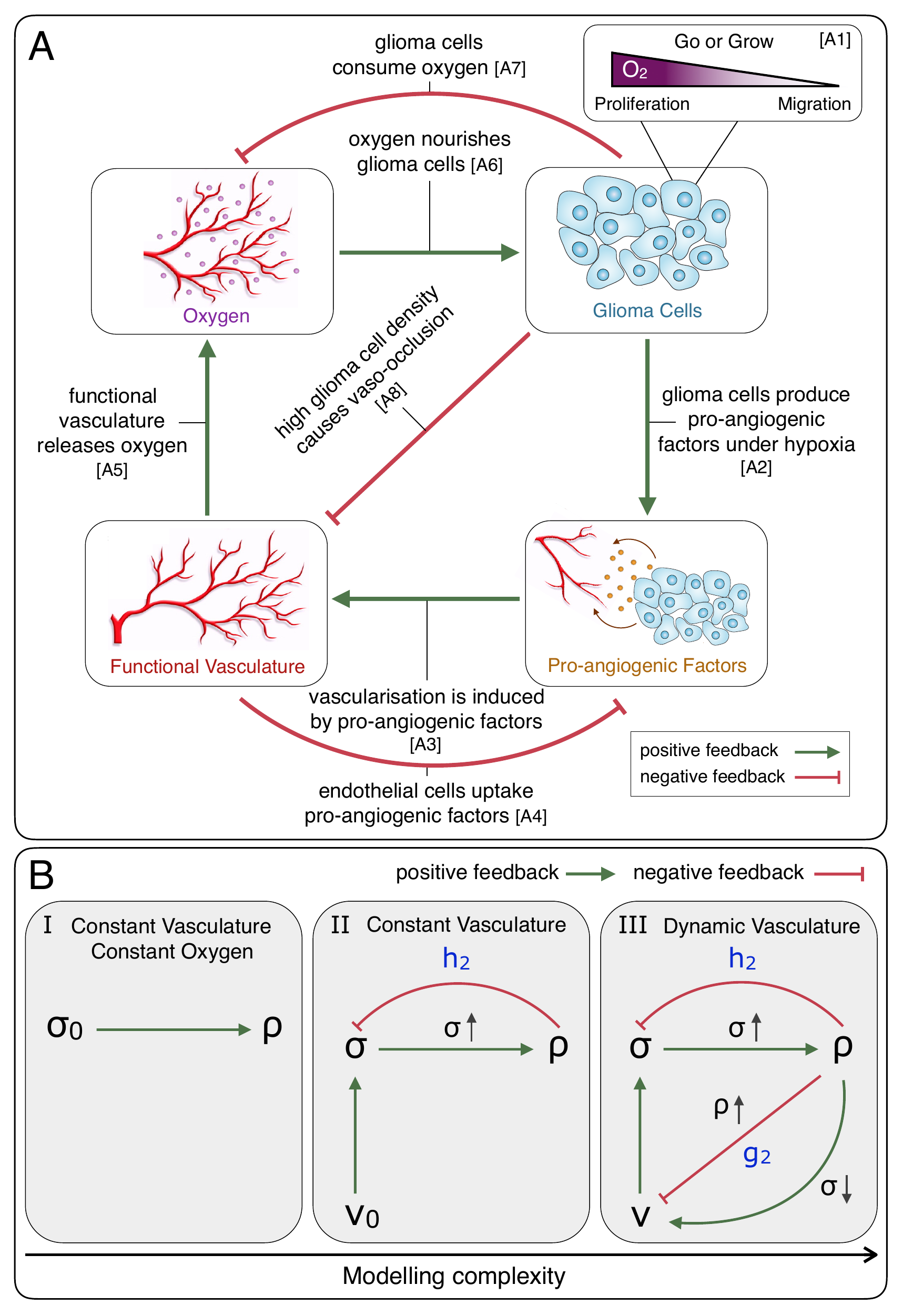}
\caption{\textbf{Modelling logic and hierarchy}. (A) Diagram of the interactions between glioma cells, oxygen, functional tumour-associated vasculature and pro-angiogenic factors. (B) From left to right model complexity increases with respect to the interactions between system variables: density of glioma cells $\rho$, density of functional tumour vasculature $v$ and oxygen concentration $\sigma$. $\sigma_0$ and $v_0$ represent a constant oxygen concentration and functional tumour vascularisation. The model parameters $g_2$ and $h_2$ are the rates of vaso-occlusion and glioma cell oxygen consumption, respectively (see equations (\ref{eq15})-(\ref{eq16})).}
\label{fig2}
\end{figure}

The mathematical model we develop describes the growth of vascularised gliomas focusing on the interplay between the migration/proliferation dichotomy and vaso-occlusion at the margin of viable tumour tissue. The system variables are density of glioma cells $\rho(x,t)$ and functional tumour vasculature $v(x,t)$, as well as concentrations of oxygen $\sigma(x,t)$ and pro-angiogenic factors $a(x,t)$ in the tumour microenvironment, where $(x,t)\in \mathbb{R}^d\times \mathbb{R}$ and $d$ is the dimension of the system. Figure~\ref{fig2}(A) shows a diagram of the system interactions/assumptions considered, which are summarised as follows: \\

\noindent {\bf [A1]} Glioma cells switch phenotypes between proliferative (normoxic) and migratory (hypoxic) depending on the oxygen concentration in the tumour microenvironment \cite{Giese1996, Giese2003, Hatzikirou2012, Bottger2012, Bottger2015}. \\
\noindent {\bf [A2]} Under hypoxia conditions glioma cells secrete large amounts of pro-angiogenic factors \cite{Carmeliet2011, Weis2011, Jain2007, Jain2014a}. \\
\noindent {\bf [A3]} Pro-angiogenic factors drive new blood vessel formation and vasculature remodelling \cite{Weis2011, Jain2013}. \\
\noindent {\bf [A4]} Endothelial cells uptake pro-angiogenic factors \cite{Weis2011, Nakayama2013}. \\
\noindent {\bf [A5]} Functional tumour-associated vasculature releases oxygen \cite{Jain2007, Carmeliet2011, Weis2011}. \\
\noindent {\bf [A6]} Oxygen availability is essential for glioma growth and progression \cite{Carmeliet2011, Weis2011, Jain2014b}. \\
\noindent {\bf [A7]} Glioma cells consume oxygen provided by the existing functional vascular network \cite{Carmeliet2011, Rockne2015}. \\
\noindent {\bf [A8]} Prothrombotic factors and high mechanical pressure induce vaso-occlusion in gliomas \cite{Brat2004b, Padera2004, Rong2006, Jain2014b}.

%--------------------------------------------------------------------------------------------------------------------------------------------------------------------------------------------------------------------------------------------------------------------%
%--------------------------------------------------------------------------------------------------------------------------------------------------------------------------------------------------------------------------------------------------------------------%
%--------------------------------------------------------------------------------------------------------------------------------------------------------------------------------------------------------------------------------------------------------------------%

\subsubsection*{Density of glioma cells}

Based on the migration/proliferation dichotomy \cite{Giese1996, Giese2003, Hatzikirou2012, Bottger2012, Bottger2015}, we assume that glioma cells $\rho(x,t)$ switch between two different cell phenotypes, migratory $\rho_1(x,t)$ (hypoxic) and proliferative $\rho_2(x,t)$ (normoxic), depending on the concentration of oxygen in the tumour microenvironment described by $\sigma(x,t)$. More precisely, we consider two linear switching functions, $f_{21}(\sigma) = \lambda_{1} - \sigma$ and $f_{12}(\sigma) = \lambda_{2} \sigma$, that represent the rate at which glioma cells change from migratory to proliferative and vice versa, respectively. The parameters $\lambda_{1}$ and $\lambda_{2}$ are positive constants, see the Supplementary Material for further details. Cell motility is modelled as a diffusive process mimicking the net invasion of glioma cells into the surrounding brain tissue, while a logistic growth term is considered for tumour cell proliferation. Accordingly, the system of equations governing the dynamics of migratory and proliferative glioma cells is given by

\begin{eqnarray}
\label{eq1} \frac{\partial \rho_{1}}{\partial t} &=& D_{\rho} \nabla^{2}\rho_{1} - f_{12}(\sigma)\rho_{1} + f_{21}(\sigma)\rho_{2}, \\
\label{eq2} \frac{\partial \rho_{2}}{\partial t} &=& b_{\rho} \hspace{0.5mm} \rho_{2}\left( 1- (\rho_{1}+\rho_{2})/N \right) + f_{12}(\sigma)\rho_{1} - f_{21}(\sigma)\rho_{2},
\end{eqnarray}

\noindent where the temporal $t$ and spatial $x$ coordinates in the arguments of variables have been omitted for notational simplicity. $D_{\rho}$ and $b_{\rho}$ are the diffusion and proliferation rates of migratory and proliferative glioma cells, respectively. $N$ represents the brain tissue carrying capacity, i.e. the maximum number of cells that can be located within a domain element. The model parameters $D_{\rho}$, $b_{\rho}$ and $N$ are positive constants.
\vspace{2mm}

The system (\ref{eq1})-(\ref{eq2}) is reduced to a single equation for the total density of glioma cells $\rho = \rho_{1} + \rho_{2}$ by assuming that $f_{12}(\sigma)\rho_{1} = f_{21}(\sigma)\rho_{2}$. This assumption implies that each phenotypic switching event is faster compared to migration and proliferation cell processes, which allows to express $\rho_{1}$ and $\rho_2$ as a function of $\rho$ in the following form

\begin{equation*}
\label{eq3} \rho = \left( 1 + \frac{f_{12}(\sigma)}{f_{21}(\sigma)} \right) \rho_{1} = \left( 1 + \frac{f_{21}(\sigma)}{f_{12}(\sigma)} \right) \rho_{2},
\end{equation*}

\noindent where we have that

\begin{equation*}
\label{eq4} \rho_{1} = \left( \frac{1}{1 + f_{12}(\sigma) / f_{21}(\sigma)} \right) \rho
\end{equation*}

\noindent and

\begin{equation*}
\label{eq5} \rho_{2} = \left( \frac{1}{1 + f_{21}(\sigma) / f_{12}(\sigma)} \right) \rho.
\end{equation*}

Summing equations (\ref{eq1}) and (\ref{eq2}), and substituting the expressions above for $\rho_{1}$ and $\rho_2$, we obtain the governing equation for the total (migratory and proliferative) density of glioma cells as follows

\begin{equation}
\label{eq6} \frac{\partial \rho}{\partial t} = D_{\rho} \nabla^{2}(\alpha(\sigma)\rho) + b_{\rho} \hspace{0.5mm} \beta(\sigma)\rho\left(1-(\alpha(\sigma) + \beta(\sigma)) \rho / N \right),
\end{equation}

\noindent where the oxygen-dependent functions $\alpha(\sigma)$ and $\beta(\sigma)$ are given by

\begin{equation}
\label{eq7} \alpha(\sigma) = \frac{1}{1+ f_{12}(\sigma) / f_{21}(\sigma)} = \frac{\lambda_{1} - \sigma}{(\lambda_{2} - 1) \sigma + \lambda_{1}},
\end{equation}

\noindent and

\begin{equation}
\label{eq8} \beta(\sigma) = \frac{1}{1+ f_{21}(\sigma) / f_{12}(\sigma)} = \frac{\lambda_{2} \sigma}{(\lambda_{2} - 1) \sigma + \lambda_{1}}.
\end{equation}

Then, taking into account that $\alpha(\sigma) + \beta(\sigma)=1$, we can rewrite equation (\ref{eq6}) as

\begin{equation}
\label{eq9} \frac{\partial \rho}{\partial t} = D_{\rho} \nabla^{2}(\alpha(\sigma)\rho) + b_{\rho} \hspace{0.5mm} \beta(\sigma) \rho \left( 1- \rho / N \right).
\end{equation}

We notice that equation (\ref{eq9}) is a generalisation of the widely studied Fisher-Kolmogorov model to describe glioma invasion \cite{Murray2002, Harpold2007}. The nonlinear terms $\alpha(\sigma)$ and $\beta(\sigma)$ in equation (\ref{eq9}) modify the rates of cell diffusion and proliferation according to oxygen availability. Under hypoxic conditions cell diffusion increases, while proliferation decreases, i.e. glioma cells become more migratory and less proliferative. On the contrary, for normal oxygen levels glioma cells become more proliferative and less invasive. Let $\sigma{_0} > 0$ be the physiological concentration of oxygen in the host brain tissue. Then, by normalising $D_{\rho} = D/\alpha(\sigma_{0})$ and $b_{\rho} = b/\beta(\sigma_{0})$ the classical Fisher-Kolmogorov equation is recovered under the assumption of a constant oxygen concentration

\begin{equation}
\label{eq10} \frac{\partial \rho}{\partial t} = D \nabla^{2}\rho + b \hspace{0.5mm} \rho \left( 1- \rho / N \right),
\end{equation}

\noindent where $D$ and $b$ are positive constants denoting respectively the intrinsic rates of diffusion and proliferation of glioma cells. Equation (\ref{eq10}) has been extensively used to predict untreated glioma invasion kinetics, as well as to estimate patient-specific parameters based on standard medical imaging \cite{Swanson2002b, Harpold2007, Swanson2011, Hawkins2013}. Furthermore, this model allowed for suitable estimations of glioma recurrence after surgical resection \cite{Swanson2003} and simulations of tumour responses to conventional therapeutic modalities as chemo- \cite{Swanson2002a} and radiation therapy \cite{Rockne2010}.

%--------------------------------------------------------------------------------------------------------------------------------------------------------------------------------------------------------------------------------------------------------------------%
%--------------------------------------------------------------------------------------------------------------------------------------------------------------------------------------------------------------------------------------------------------------------%
%--------------------------------------------------------------------------------------------------------------------------------------------------------------------------------------------------------------------------------------------------------------------%

\subsubsection*{Pro-angiogenic factor concentration}

Neovascularisation in tumours takes place when pro-angiogenic factors overcome anti-angiogenic stimuli. However, in gliomas there is a wide range of pro- and anti-angiogenic factors involved, each of them acting through different vascularisation mechanisms \cite{Jain2007, Jain2013, Jain2014a}. While not explicitly considering the vascular endothelial growth factor (VEGF) or any other pro-angiogenic molecule, we assume a generic effective pro-angiogenic factor concentration $a$ at quasi-steady state. In fact, we suppose that an over-expression of pro-angiogenic factors instantaneously promotes formation of functional tumour vasculature. We further assume that pro-angiogenic factors are only produced by glioma cells under hypoxic conditions at a rate proportional to tumour cell density, and therefore neglect hypoxia-independent pathways. Moreover, pro-angiogenic factors are consumed by endothelial cells and undergo natural decay. The equation for the effective pro-angiogenic factor concentration $a(x,t)$ is given by

\begin{equation*}
\label{eq11} 0 = k_{\text{1}} \hspace{0.5mm} \rho \hspace{0.5mm} \tilde{\text{H}}_{\theta}(\sigma-\sigma_a^{*}) - k_{2}av  - k_{3}a,
\end{equation*}

\noindent where

\begin{equation}
\label{eq12} a = \frac{k_{1} \hspace{0.5mm} \rho \hspace{0.5mm} \tilde{\text{H}}_{\theta}(\sigma - \sigma_a^{*})}{k_{2}v+k_{3}}.
\end{equation}

The positive constants $k_{1}$, $k_{2}$ and $k_{3}$ represent the production, consumption and natural decay rates, respectively, where $0 < \sigma_a^{*} < \sigma_0$ is the hypoxic oxygen threshold for production of pro-angiogenic factors by glioma cells. The function $\tilde{\text{H}}_{\theta}(\sigma - \sigma_a^{*})$ is a continuous approximation of the Heaviside decreasing step function $H(\xi)$, which is defined as $H(\xi) = 1$ if $\xi \leq 0$ and $H(\xi) = 0$ if $\xi > 0$, and given by

\begin{equation}
\label{eq13} \tilde{\text{H}}_{\theta}(\sigma - \sigma_a^{*})=1-\frac{1}{1+e^{-2\theta (\sigma - \sigma_a^{*})}},
\end{equation}

\noindent where $\theta$ is a positive constant that controls the steepness of $\tilde{\text{H}}_{\theta}$ at $(\sigma - \sigma_a^{*})$.

%--------------------------------------------------------------------------------------------------------------------------------------------------------------------------------------------------------------------------------------------------------------------%
%--------------------------------------------------------------------------------------------------------------------------------------------------------------------------------------------------------------------------------------------------------------------%
%--------------------------------------------------------------------------------------------------------------------------------------------------------------------------------------------------------------------------------------------------------------------%

\subsubsection*{Density of functional tumour vasculature}

Several experimental findings support that vascular structure and function become markedly abnormal in brain tumours \cite{Carmeliet2011, Weis2011, Jain2014b}. Malignant gliomas, and particularly glioblastomas, have blood vessels of increased diameter, high permeability, thickened basement membranes and highly proliferative endothelial cells \cite{Jain2007}, see also Figure~\ref{fig1}(B). Due to such abnormalities, a significant fraction of the tumour-associated vascular network does not constitute functional blood vessels \cite{Jain2007}. Based on these observations, we exclusively account for functional vascularisation instead of modelling the total density of tumour blood vessels. Accordingly, we assume that the density of functional tumour vasculature $v(x,t)$ is a dimensionless and normalised quantity with values in the interval $[0, 1]$. The normal density of functional vascularisation in the host brain tissue is taken equal to $v = 1/2$. The limit case $v = 0$ represents an avascular tissue, while on the contrary $v = 1$ describes a hypothetical scenario characterised by excessive vascularisation.
\vspace{2mm}

Blood vessels in gliomas are not stable, being continuously formed, occluded and destroyed. Neovascularisation takes place by different angiogenic and vasculogenic processes induced by complex signalling mechanisms that are not well understood \cite{Hardee2012, Kohn2013, Sugihara2015}. For simplicity, we assume that tumour blood vessels are created when pro-angiogenic factors prevail anti-angiogenic stimuli, i.e. for $a > 0$, leading to development of new functional vasculature according to a logistic growth term. The rate at which such vasculature is generated follows the Michaelis-Menten kinetics depending on the pro-angiogenic factor concentration, where diffusive vascular dispersal at a constant rate is assumed. On the other hand, mechanical or chemical cues in regions of high glioma cell density induce blood vessel occlusion or collapse \cite{Brat2004b, Padera2004, Rong2006}. Vaso-occlusion is then modelled by an exponential term depending on the density of glioma cells. The equation for the density of functional tumour vasculature $v(x,t)$ is given by

\begin{equation}
\label{eq14} \frac{\partial v}{\partial t} = D_{v}\nabla^{2}v + g_{1}\frac{a}{\mu+a}v\left(1-v\right) - g_{2} v \rho^{n},
\end{equation}

\noindent where the temporal $t$ and spatial $x$ coordinates in the arguments of variables have been omitted for notational simplicity. $D_{v}$ is the diffusion coefficient representing the net dispersal of functional tumour vasculature, $g_{1}$ is the maximum formation rate of functional blood vessels, $\mu$ is the pro-angiogenic factor concentration at which $g_{1}$ is half-maximal, $g_{2}$ is the vaso-occlusion rate and $n$ is a parameter that controls the degree of vaso-occlusion depending on the density of glioma cells. The model parameters $D_{v}$, $g_{1}$, $\mu$, $g_{2}$ and $n$ are positive constants.
\vspace{2mm}

Plugging equation (\ref{eq12}) for effective pro-angiogenic factor concentration into equation (\ref{eq14}), and assuming that the decay rate of $a$ is much smaller than the consumption rate by endothelial cells, i.e. $k_3 \ll k_2$ \cite{KohnLuque2013}, we obtain that

\begin{equation}
\label{eq15} \frac{\partial v}{\partial t}  = D_{v}\nabla^{2}v + g_{1}\dfrac{ \dfrac{\rho}{v}\tilde{\text{H}}_{\theta}(\sigma-\sigma^{*}_{a})} {K + \dfrac{\rho}{v}\tilde{\text{H}}_{\theta}(\sigma-\sigma^{*}_{a})}v\left(1-v\right) - g_{2} v \rho^{n},
\end{equation}

\noindent where $K = \mu k_{2} / k_{1}$ is a positive constant denoting the concentration of pro-angiogenic factors at which the functional tumour vasculature formation rate is half-maximal, see the Supplementary Material for more details.

%--------------------------------------------------------------------------------------------------------------------------------------------------------------------------------------------------------------------------------------------------------------------%
%--------------------------------------------------------------------------------------------------------------------------------------------------------------------------------------------------------------------------------------------------------------------%
%--------------------------------------------------------------------------------------------------------------------------------------------------------------------------------------------------------------------------------------------------------------------%

\subsubsection*{Oxygen concentration}

Oxygen is delivered to the host brain tissue via functional blood vessels, spreads into the tumour mass and is consumed by glioma cells. Transport of oxygen within tissues occurs by diffusion and convection \cite{Jain1999}. For simplicity, we neglect the convective contributions and only consider that after transvascular exchange oxygen molecules move exclusively by diffusion. Oxygen supply is modelled by assuming that the supply rate is proportional to the functional vascularisation and the difference between the physiological oxygen concentration in the host brain tissue and that in the tumour interstitium. These assumptions result in the following equation for the oxygen concentration $\sigma(x,t)$

\begin{equation}
\label{eq16} \frac{\partial \sigma}{\partial t} = D_{\sigma}\nabla^{2}\sigma + h_{1} v \left(\sigma_{0} - \sigma \right) -  h_{2}\rho \sigma,
\end{equation}

\noindent where the temporal $t$ and spatial $x$ coordinates in the arguments of variables have been omitted for notational simplicity. $D_{\sigma}$ is the oxygen diffusion coefficient, $h_{1}$ is the permeability coefficient of functional blood vessels, $\sigma_{0}$ is the physiological oxygen concentration in the host brain tissue and $h_{2}$ is the oxygen consumption rate by glioma cells. The model parameters $D_{\sigma}$, $h_{1}$, $\sigma_{0}$ and $h_{2}$ are positive constants. Similar assumptions have been previously considered to model oxygen-related mechanisms in tumour growth \cite{Stamper2010}.

%--------------------------------------------------------------------------------------------------------------------------------------------------------------------------------------------------------------------------------------------------------------------%
%--------------------------------------------------------------------------------------------------------------------------------------------------------------------------------------------------------------------------------------------------------------------%
%--------------------------------------------------------------------------------------------------------------------------------------------------------------------------------------------------------------------------------------------------------------------%

\subsubsection*{Model formulation, boundary and initial conditions}

The proposed model of glioma-vasculature interplay comprises the following system of coupled partial differential equations

\begin{eqnarray}
\label{eq17} \frac{\partial \rho}{\partial t} &=& D_{\rho} \nabla^{2}(\alpha(\sigma)\rho) + b_{\rho} \hspace{0.5mm} \beta(\sigma)\rho\left(1- \rho/N \right), \\
\label{eq18} \frac{\partial v}{\partial t} &=& D_{v}\nabla^{2}v + g_{1}\dfrac{ \dfrac{\rho}{v}\tilde{\text{H}}_{\theta}(\sigma-\sigma^{*}_{a})} {K + \dfrac{\rho}{v}\tilde{\text{H}}_{\theta}(\sigma-\sigma^{*}_{a})}v\left(1-v\right) - g_{2} v \rho^{n}, \\
\label{eq19} \frac{\partial \sigma}{\partial t} &=& D_{\sigma}\nabla^{2}\sigma + h_{1}v \left(\sigma_{0} - \sigma \right) -  h_{2} \rho \sigma,
\end{eqnarray}

\noindent where the oxygen-dependent functions $\alpha(\sigma)$ and $\beta(\sigma)$ are given by equations (\ref{eq7})-(\ref{eq8}), respectively. The system of equations above is closed by imposing the following initial conditions

\begin{eqnarray*}
\rho(x,0)      &=& \rho_{0} \tilde{H}_{\gamma}(x-\epsilon) = \rho_{0} \left( 1 - \frac{1}{1+e^{-2 \gamma (x-\epsilon)}} \right), \hspace{5mm} 0 \leq x \leq L, \\
v(x,0)           &=& v_{0}, \hspace{62mm} 0 \leq x \leq L, \\
\sigma(x,0)  &=& \sigma_{0}, \hspace{62mm} 0 \leq x \leq L,
\end{eqnarray*}

\noindent where the positive constants $\rho_{0}$, $\sigma_{0}$ and $v_{0}$ are the initial density of glioma cells located in a small segment of length $\epsilon$, density of functional tumour vasculature and oxygen concentration, respectively. The length of the one-dimensional simulation domain is represented by $L > 0$, and $\gamma$ is a positive constant that controls the steepness of $\tilde{H}_{\gamma}$ at $(x-\epsilon)$ with $\epsilon > 0$. Moreover, we consider an isolated host tissue in which all behaviours arise due to the interaction terms. This assumption results in no-flux boundary conditions of the form

\begin{eqnarray*}
\begin{split}
\rho_{x}(0,t) &=& v_{x}(0,t) &=& \sigma_{x}(0,t) &=& 0, \hspace{5mm} 0 \leq t \leq T_{f}, \\
\rho_{x}(L,t)  &=& v_{x}(L,t)  &=&\sigma_{x}(L,t)   &=& 0,  \hspace{5mm} 0 \leq t \leq T_{f},
\end{split}
\end{eqnarray*}

\noindent where $T_{f} > 0$ is an arbitrary time. These boundary conditions also imply that no cell or molecule leaves the system through the tissue/domain boundaries.

%--------------------------------------------------------------------------------------------------------------------------------------------------------------------------------------------------------------------------------------------------------------------%
%--------------------------------------------------------------------------------------------------------------------------------------------------------------------------------------------------------------------------------------------------------------------%
%--------------------------------------------------------------------------------------------------------------------------------------------------------------------------------------------------------------------------------------------------------------------%

\subsection*{Modelling hierarchy}

The glioma-vasculature interplay model given by equations (\ref{eq17})-(\ref{eq19}), and referred to as {\it model~III}, is a generalisation of two simpler models which are also of interest for the study of glioma invasion. As shown in Figure~\ref{fig2}(B), such simpler models are obtained under the assumptions of constant density of functional tumour vasculature $v(x,t) = v_0$ ({\it model~II}), and also constant oxygen concentration $\sigma(x,t) = \sigma_{0}$ ({\it model~I}). Specifically, {\it model~II} is obtained from {\it model~III} by setting $g_{1} = g_{2} = 0$ in equation (\ref{eq18}), i.e. assuming neither formation nor occlusion/collapse of tumour blood vessels. In turn, {\it model~I} is obtained from {\it model~II} by setting $h_{2} = 0$ in equation (\ref{eq19}), i.e. assuming a constant concentration of oxygen in the tumour microenvironment.
\vspace{2mm}

{\it Model~I} is similar to the classical Fisher-Kolmogorov equation (\ref{eq10}), for which a large number of theoretical and simulation results are known \cite{Murray2002, Harpold2007}. {\it Model~II} given by equations (\ref{eq17}) and (\ref{eq19}) contains an extended version of the Fisher-Kolmogorov equation with nonlinear glioma cell diffusion and proliferation terms. Both nonlinearities depend on the oxygen concentration in the tumour microenvironment, which is governed by a reaction-diffusion equation with linear diffusion and nonlinear reaction terms. In addition, the dynamics of the glioma cell population are modelled by considering the migration/proliferation dichotomy (Go-or-Grow). As in {\it model~II} the supply of oxygen is assumed constant, the blood perfusion is stable and we neglect tumour-induced vascular pathologies. The latter is a a reasonable assumption, especially for low grade gliomas, where abnormal vasculature is not prominent \cite{Swanson2011}. A natural extension of {\it model~II} is to consider tumour-associated vascularisation dynamics. Accordingly, {\it model~III} is formulated to investigate the effects of vaso-modulatory interventions on glioma invasion. Taking into account the huge amount of results reported from {\it model~I}, we analyse {\it model~II} as an intermediate step towards the study of {\it model~III}, see Figure~\ref{fig2}(B). In particular, we focus on the impact of glioma cell oxygen consumption and vaso-occlusion modulations on tumour front speed and infiltration width. In the Supplementary Material we provide details about model simulations and the numerical implementation.

%--------------------------------------------------------------------------------------------------------------------------------------------------------------------------------------------------------------------------------------------------------------------%
%--------------------------------------------------------------------------------------------------------------------------------------------------------------------------------------------------------------------------------------------------------------------%
%--------------------------------------------------------------------------------------------------------------------------------------------------------------------------------------------------------------------------------------------------------------------%

\subsection*{Model observables}

We characterise glioma invasion by the tumour front speed and infiltration width, see Figure~S1 in the Supplementary Material. The tumour front speed is estimated by the rate of change given by the point of maximum slope in $\rho(x,t)$ at the end of numerical simulations $T_f$. In turn, the infiltration width is defined by the difference between the points where glioma cell density is $80\%$ and $2\%$ of the maximum cell density at time $T_f$. These tumour invasion properties have been reported crucial to determine glioma malignancy and therapeutic failure rates \cite{Swanson2003, Harpold2007, Swanson2011}.
\vspace{2mm}

Unlike the mathematical model given by the classical Fisher-Kolmogorov equation (\ref{eq10}), in our glioma invasion model given by equations (\ref{eq17})-(\ref{eq19}) cell processes are regulated by oxygen availability. Thus, we distinguish intrinsic cell diffusion $D$ and proliferation $b$ rates from effective rates which take into account the oxygen concentration in the tumour microenvironment. Accordingly, the effective diffusion $D_{\mbox{eff}}$ and proliferation $b_{\mbox{eff}}$ rates are defined as

\begin{equation}
\label{eq20} D_{\mbox{eff}} = D_{\rho} \hspace{1mm} L^{-1} \int\limits_{L} \alpha(\sigma(x,t)) \hspace{1mm} dx
\end{equation}

\noindent and

\begin{equation}
\label{eq21} b_{\mbox{eff}} = b_{\rho} \hspace{1mm} L^{-1} \int\limits_{L} \beta(\sigma(x,t)) \hspace{1mm} dx,
\end{equation}

\noindent where $L$ represents the length of the one-dimensional domain of simulation, $D_{\rho} = D/\alpha(\sigma_{0})$ and $b_{\rho} = b/\beta(\sigma_{0})$, where $D$ and $b$ are the intrinsic rates of glioma cell diffusion and proliferation, respectively. The parameter $\sigma{_0}$ is the physiological concentration of oxygen in the host brain tissue. In the following, we investigate the dependence of $D_{\mbox{eff}}$ and $b_{\mbox{eff}}$ at time $T_f$, as well as the tumour front speed and infiltration width, for different ranges of model parameters $h_2$ (cell oxygen consumption) and $g_2$ (vaso-occlusion).

%--------------------------------------------------------------------------------------------------------------------------------------------------------------------------------------------------------------------------------------------------------------------%
%--------------------------------------------------------------------------------------------------------------------------------------------------------------------------------------------------------------------------------------------------------------------%
%--------------------------------------------------------------------------------------------------------------------------------------------------------------------------------------------------------------------------------------------------------------------%

\subsection*{Model parameterisation}

Model parameter values are taken from published data wherever possible or estimated to approximate physiologic conditions based on appropriate physical arguments, see Table~\ref{tab1} and the Supplementary Material for further details. For parameters of special interest, a wide range of values is considered to explore their effects on the resulting glioma invasion.

\begin{table}[H]
\caption{Model parameter values (see the Supplementary Material).}
\label{tab1}
\footnotesize
\begin{tabular}{l l c c l}
\toprule
{\bf Parameter} & {\bf Description} & {\bf Value} & {\bf Source} \\
\hline
Glioma Cells \\
\hline
$D$               & Intrinsic diffusion rate of glioma cells                      & [$2.73 \times 10^{-3}$, $2.73 \times 10^{-1}$]~mm$^2$~day$^{-1}$  & \cite{Harpold2007, Swanson2011, Badoual2014} \\
$b$                & Intrinsic proliferation rate of glioma cells                & [$2.73 \times 10^{-4}$, $2.73 \times 10^{-2}$]~day$^{-1}$                  & \cite{Harpold2007, Swanson2011, Badoual2014} \\
$N$               & Brain tissue carrying capacity                                 & $10^{2}$~cells~mm$^{-1}$                                                                   & \cite{Eikenberry2009, McDaniel2013} \\
$\sigma_0$   & Physiological oxygen concentration                        & 1.0~nmol~mm$^{-1}$                                                                           & \cite{Hoffman1996, Carreau2011} \\
$\lambda_1$ & Phenotypic switching parameter $^{(\dagger)}$     & 2.0~nmol~mm$^{-1}$                                                                           & {\it Model specific} \\
$\lambda_2$ & Phenotypic switching parameter $^{(\ddagger)}$   & $\{0.5,~1.0,~2.0\}$                                                                                & {\it Model specific} \\
\hline
Oxygen \\
\hline
$D_{\sigma}$ & Diffusion rate of oxygen    & $1.51 \times 10^{2}$~mm$^2$~day$^{-1}$                                                & \cite{Matzavinos2009, Stamper2010, Powathil2012} \\
$h_{1}$          & Oxygen supply rate           & $3.37 \times 10^{-1}$~day$^{-1}$                                                               & \cite{Eggleton1998, Goldman2000, Kelly2006} \\
$h_{2}$          & Oxygen consumption rate & $[5.73 \times 10^{-3},~1.14 \times 10^{-1}]$~mm~cell$^{-1}$~day$^{-1}$ & \cite{Vaupel1989, Grimes2014} \\
\hline
Vasculature \\
\hline
$D_{v}$               & Vasculature dispersal rate                                     & $5.0 \times 10^{-4}$~mm$^2$~day$^{-1}$                                                             & \cite{Anderson1998, Stamper2010, Swanson2011} \\
$g_{1}$               & Vasculature formation rate                                     & $10^{-1}$~day$^{-1}$                                                                                             & \cite{Shaifer2010, Stamper2010, Scianna2013} \\
$\sigma^{*}_{a}$ & Oxygen concentration threshold for hypoxia          & $2.5 \times 10^{-1}$~nmol~mm$^{-1}$                                                                   & \cite{Cardenas2004, Vaupel2007, Powathil2012} \\
$K$                     & Half-maximal pro-angiogenic factor concentration & 1.0~nmol~mm$^{-1}$                                                                                              & {\it Estimated} \\
$g_2$                 & Vaso-occlusion rate                                                & [$5.0 \times 10^{-13}$, $1.5 \times 10^{-11}$]~cell$^{-n}$~mm$^{n}$~day$^{-1}$ & {\it Estimated} \\
$n$                     & Dimensionless vaso-occlusion degree                   & 6                                                                                                                               & {\it Estimated} \\
\bottomrule
\end{tabular}
\footnotesize{$(\dagger)$ Proliferative to migratory. $(\ddagger)$ Migratory to proliferative.}
\end{table}

%--------------------------------------------------------------------------------------------------------------------------------------------------------------------------------------------------------------------------------------------------------------------%
%--------------------------------------------------------------------------------------------------------------------------------------------------------------------------------------------------------------------------------------------------------------------%
%--------------------------------------------------------------------------------------------------------------------------------------------------------------------------------------------------------------------------------------------------------------------%

\section*{Results}

%--------------------------------------------------------------------------------------------------------------------------------------------------------------------------------------------------------------------------------------------------------------------%
%--------------------------------------------------------------------------------------------------------------------------------------------------------------------------------------------------------------------------------------------------------------------%
%--------------------------------------------------------------------------------------------------------------------------------------------------------------------------------------------------------------------------------------------------------------------%

\subsection*{Increasing cell oxygen consumption and vaso-occlusion result in more diffusive and less proliferative gliomas}

\begin{figure}[H]
\centering
\includegraphics[width=0.84\textwidth]{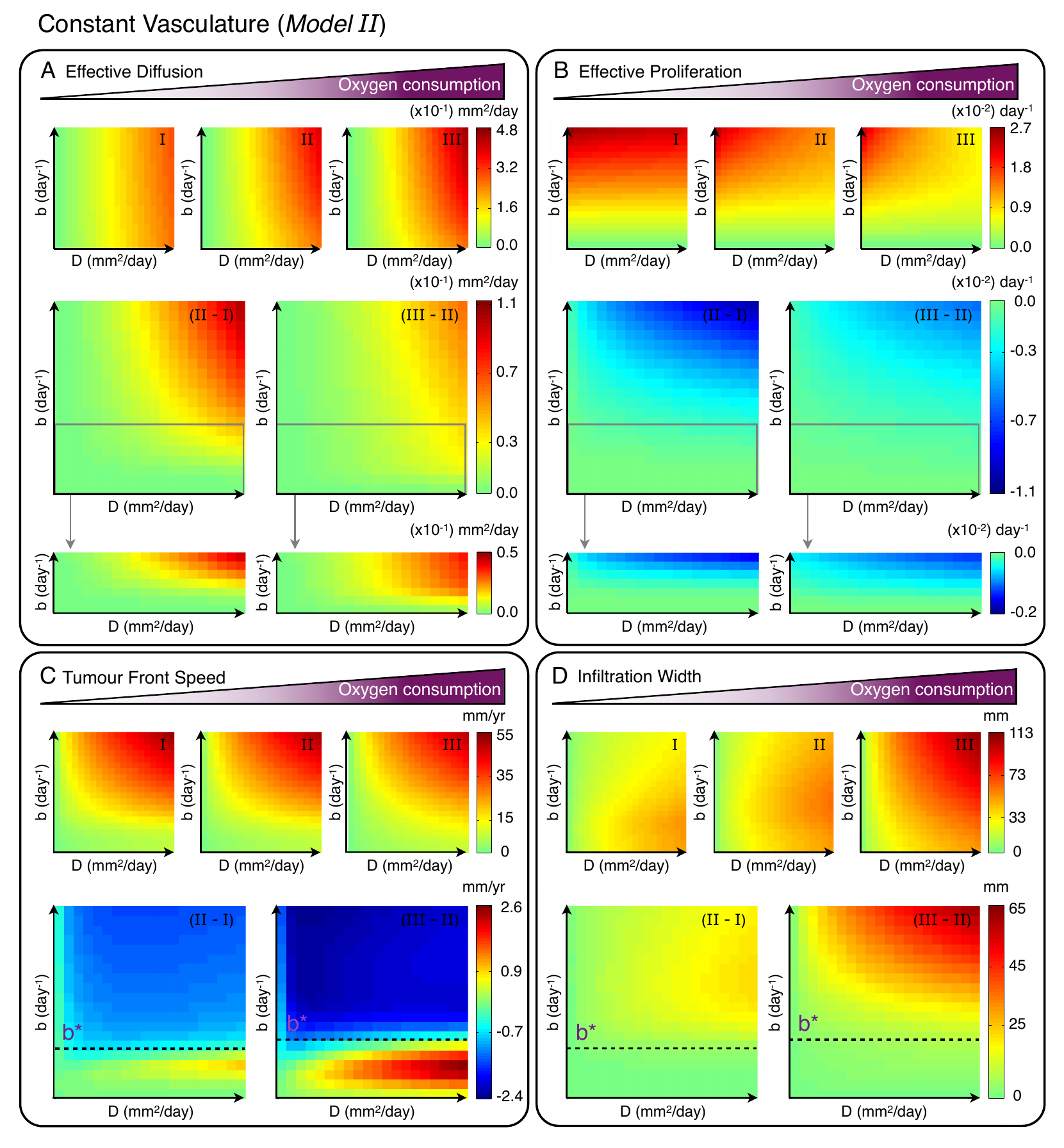}
\caption{\textbf{Oxygen consumption effects on glioma invasion for constant functional tumour vasculature.} Simulation maps with respect to the intrinsic proliferation $b \in [2.73 \times 10^{-4},~2.73 \times 10^{-2}]$~days$^{-1}$ and diffusion $D \in [2.73 \times 10^{-3},~2.73 \times 10^{-1}]$~mm$^2$~days$^{-1}$ rates of glioma cells. (A) Effective diffusion, (B) effective proliferation, (C) tumour front speed and (D) infiltration width for different oxygen consumption rates $h_2 = \{ 5.73 \times 10^{-4},~5.73 \times 10^{-3},~5.73 \times 10^{-2}\}$~mm~cell$^{-1}$~day$^{-1}$ in simulation maps I-III respectively. (A-D) Differences between simulation maps are provided. The other model parameters are as in Table~\ref{tab1}.}
\label{fig3}
\end{figure}

\begin{figure}[H]
\centering
\includegraphics[width=0.84\textwidth]{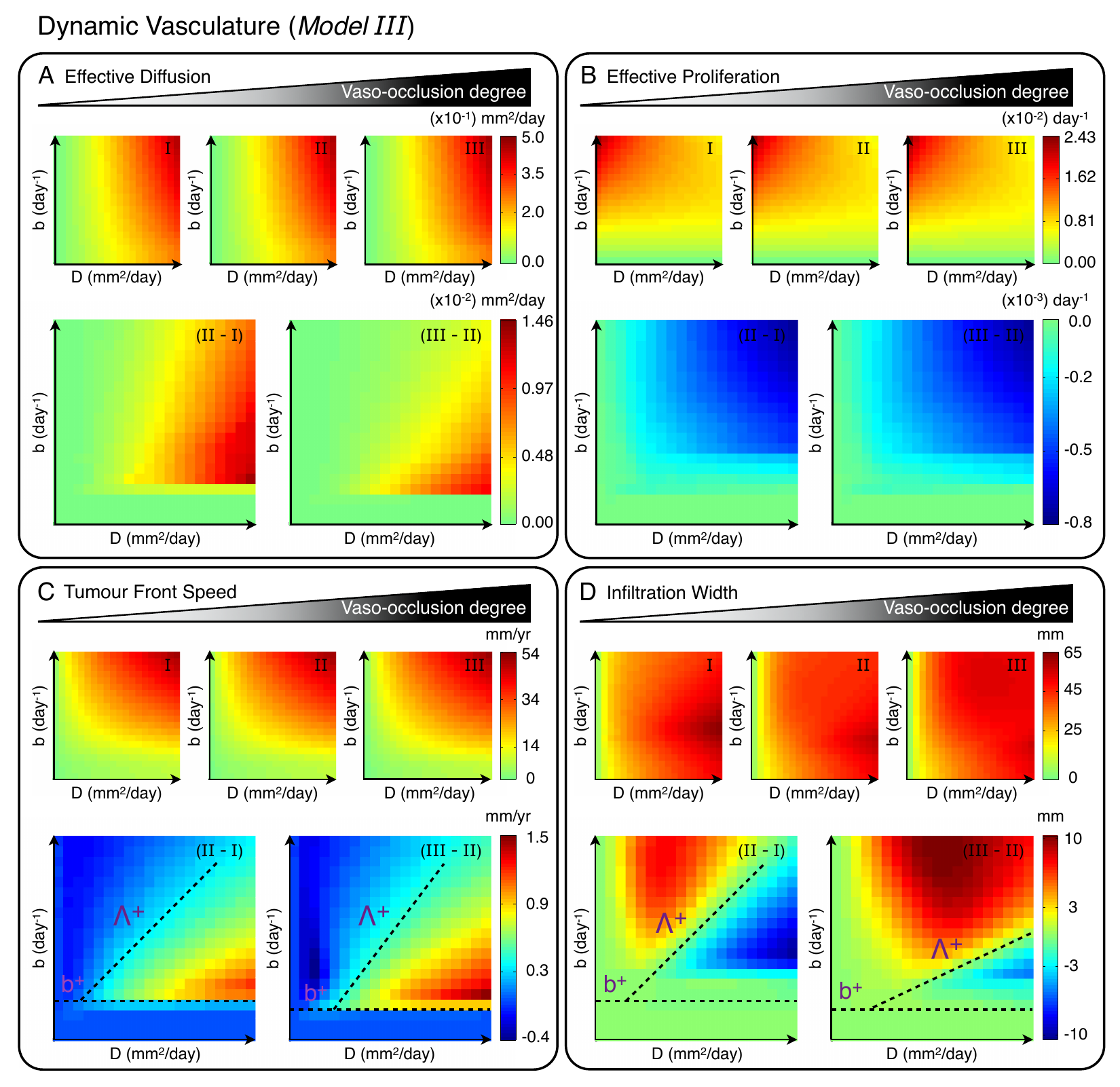}
\caption{\textbf{Vaso-occlusion effects on glioma invasion.} Simulation maps with respect to the intrinsic proliferation $b \in [2.73 \times 10^{-4},~2.73 \times 10^{-2}]$~days$^{-1}$ and diffusion $D \in [2.73 \times 10^{-3},~2.73 \times 10^{-1}]$~mm$^2$~days$^{-1}$ rates of glioma cells. (A) Effective diffusion, (B) effective proliferation, (C) tumour front speed and (D) infiltration width for a fixed oxygen consumption $h_2 = 5.73 \times 10^{-3}$~mm~cell$^{-1}$~day$^{-1}$ and different vaso-occlusion $g_2 = \{5.0 \times 10^{-13},~5.0 \times 10^{-12},~1.5 \times 10^{-11}\}$~cells$^{-n}$~mm$^{n}$~day$^{-1}$ rates in simulation maps I-III respectively. (A-D) Differences between simulation maps are provided. The other model parameters are as in Table~\ref{tab1}.}
\label{fig4}
\end{figure}

The glioma-vasculature interplay model given by equations (\ref{eq17})-(\ref{eq19}) is first used to investigate the effects of cell oxygen consumption and vaso-occlusion modulations on the effective behaviour of gliomas. Figures~\ref{fig3}(A,B) and \ref{fig4}(A,B) provide simulation maps of effective diffusion $D_{\mbox{eff}}$ and proliferation $b_{\mbox{eff}}$ rates, as defined in equations (\ref{eq20}) and (\ref{eq21}), for gliomas characterised by different combinations of intrinsic cell coefficients $D$ and $b$. Model simulations in Figure~\ref{fig3}(A,B) are obtained under the assumption of constant functional vasculature density, i.e. neither formation nor occlusion/collapse of tumour blood vessels, for increasing oxygen consumption rates by glioma cells. In turn, Figure~\ref{fig4}(A,B) provides simulation maps for a fixed oxygen consumption rate considering tumour vascularisation dynamics and increasing vaso-occlusion rates.
\vspace{2mm}

Comparative simulation maps in Figures~\ref{fig3}(A,B) and \ref{fig4}(A,B) illustrate that increasing the rate of oxygen consumption by glioma cells $h_2$ and vaso-occlusion $g_2$ result in more diffusive and less proliferative tumours. Modulations of the oxygen consumption rate have major impact on highly infiltrative and rapidly growing gliomas. At high values of both parameters, $h_2$ and $g_2$, the oxygen concentration in the tumour microenvironment significantly decreases. The lack of oxygen limits the proliferative capacity of glioma cells, and in turn enhances the hypoxia-induced cell migration towards better oxygenated brain tissue regions. The precise way in which such changes in glioma cell dynamics affect invasion responses are predicted to depend on the intrinsic tumour features.

%--------------------------------------------------------------------------------------------------------------------------------------------------------------------------------------------------------------------------------------------------------------------%
%--------------------------------------------------------------------------------------------------------------------------------------------------------------------------------------------------------------------------------------------------------------------%
%--------------------------------------------------------------------------------------------------------------------------------------------------------------------------------------------------------------------------------------------------------------------%

\subsection*{Modulations of cell oxygen consumption and vaso-occlusion rate result in opposing effects on glioma invasion}

Figures~\ref{fig3}(C,D) and \ref{fig4}(C,D) show simulation maps of tumour front speed and infiltration width with respect to different combinations of intrinsic cell coefficients $D$ and $b$. In particular, these glioma invasion properties are determined by a non-linear relationship between effective diffusion $D_{\mbox{eff}}$ and proliferation $b_{\mbox{eff}}$ of glioma cells. For instance, in the simplest case of {\it model~I} similar to the classical Fisher-Kolmogorov equation (\ref{eq10}), the tumour front speed is proportional to $\sqrt{D_{\mbox{eff}}\hspace{1.0mm}b_{\mbox{eff}}}$~and infiltration width $\sqrt{D_{\mbox{eff}} / b_{\mbox{eff}}}$. Model simulations predict that, depending on the intrinsic tumour features, modulations of cell oxygen consumption and vaso-occlusion rates produce opposing effects on the resulting front speed and infiltration width. These findings are counter-intuitive and might have important implications for possible modulatory interventions targeting cell oxygen consumption and vaso-occlusion in gliomas, as discussed below.

%--------------------------------------------------------------------------------------------------------------------------------------------------------------------------------------------------------------------------------------------------------------------%
%--------------------------------------------------------------------------------------------------------------------------------------------------------------------------------------------------------------------------------------------------------------------%
%--------------------------------------------------------------------------------------------------------------------------------------------------------------------------------------------------------------------------------------------------------------------%

\subsection*{Cell oxygen consumption variations reveal a critical proliferation rate for glioma invasion}

Model analysis, under the assumption of constant density of functional tumour vasculature, reveals that modulations of the rate $h_2$ at which glioma cells consume oxygen produce opposing effects on the tumour front speed. More precisely, Figure~\ref{fig3}(C) reveals that there exists a critical proliferation rate $b^{*}$ for which the front speed of gliomas characterised by $b > b^{*}$ decreases at higher values of $h_2$, while on the contrary tumours with $b < b^{*}$ invade faster. Assuming that tumour front speed is proportional to the product of effective diffusion and proliferation rates, we can easily understand the afore-mentioned results for variations of $h_2$. In particular, above the critical proliferation rate $b^{*}$ effective diffusion and proliferation negate each other and leave the resulting front speed almost invariant. For $b < b^{*}$, the effective tumour proliferation remains intact, but the effective diffusion capacity increases for raising $h_2$ values inducing higher front speeds.
\vspace{2mm}

The flatness/steepness of tumour fronts is determined by a relation dependent on the ratio of effective diffusion and proliferation rates. When oxygen is not limited, highly diffusive tumours evolve with large and flat fronts, whereas increased cell proliferation results in short and steep fronts. However, under oxygen-limiting conditions this relation is markedly influenced by the specific rate at which glioma cells consume oxygen. Figure~\ref{fig3}(D) shows that variations in the cell oxygen consumption rate have always the same overall impact on the tumour infiltration width. Comparative simulation maps reveal that whatever the intrinsic tumour features, an arbitrary increase (decrease) in the cell oxygen consumption rate produces larger (smaller) infiltrative responses. Indeed, the effective glioma  proliferation capacity is reduced for increasing oxygen consumption rates and in turn hypoxia-induced effective migration is enhanced, yielding more infiltrative tumour growth patterns.

%--------------------------------------------------------------------------------------------------------------------------------------------------------------------------------------------------------------------------------------------------------------------%
%--------------------------------------------------------------------------------------------------------------------------------------------------------------------------------------------------------------------------------------------------------------------%
%--------------------------------------------------------------------------------------------------------------------------------------------------------------------------------------------------------------------------------------------------------------------%

\subsection*{Modulation of tumour vaso-occlusion reveals a critical cell proliferation/diffusion ratio for glioma invasion}

Model simulations show that for rising vaso-occlusion rates $g_2$, the front speed is affected differently depending on the intrinsic diffusion and proliferation rates of glioma cells. In this case, glioma invasion is additionally influenced by vascularisation mechanisms. Comparative simulation maps in Figure~\ref{fig4}(C) suggest that tumours with features inside a region delimited by a critical proliferation rate $b^{+}$ and an approximate ratio between cell diffusion and proliferation rates $\Lambda^{+} = b/D$ invade faster as $g_2$ increases. The tumour front speed out of such region decreases or remains invariant. Gliomas characterised by $b < b^{+}$ evolve at low cell density and thus vaso-occlusive events hardly occur. On the other hand, increasing vaso-occlusion rates for $b > b^{+}$ enhances effective migration towards better vascularised brain tissue areas. Although vaso-occlusion limits the proliferative activity of glioma cells, faster front speeds are obtained as long as the induced migratory responses dominate.
\vspace{2mm}

The infiltration width of gliomas with $b < b^{+}$ is almost unaffected for increasing vaso-occlusion rates as shown in Figure~\ref{fig4}(D). However, gliomas characterised by $b > b^{+}$ are also separated by an approximated linear relationship between cell coefficients $D$ and $b$ with respect to variations in the infiltration width. In particular, increasing vaso-occlusive events results in larger flat fronts for gliomas with cell proliferation/diffusion ratios above the critical one, while the infiltration width decreases in the remaining cases.

%--------------------------------------------------------------------------------------------------------------------------------------------------------------------------------------------------------------------------------------------------------------------%
%--------------------------------------------------------------------------------------------------------------------------------------------------------------------------------------------------------------------------------------------------------------------%
%--------------------------------------------------------------------------------------------------------------------------------------------------------------------------------------------------------------------------------------------------------------------%

\section*{Discussion}

\begin{figure}[H]
\centering
\includegraphics[width=0.75\textwidth]{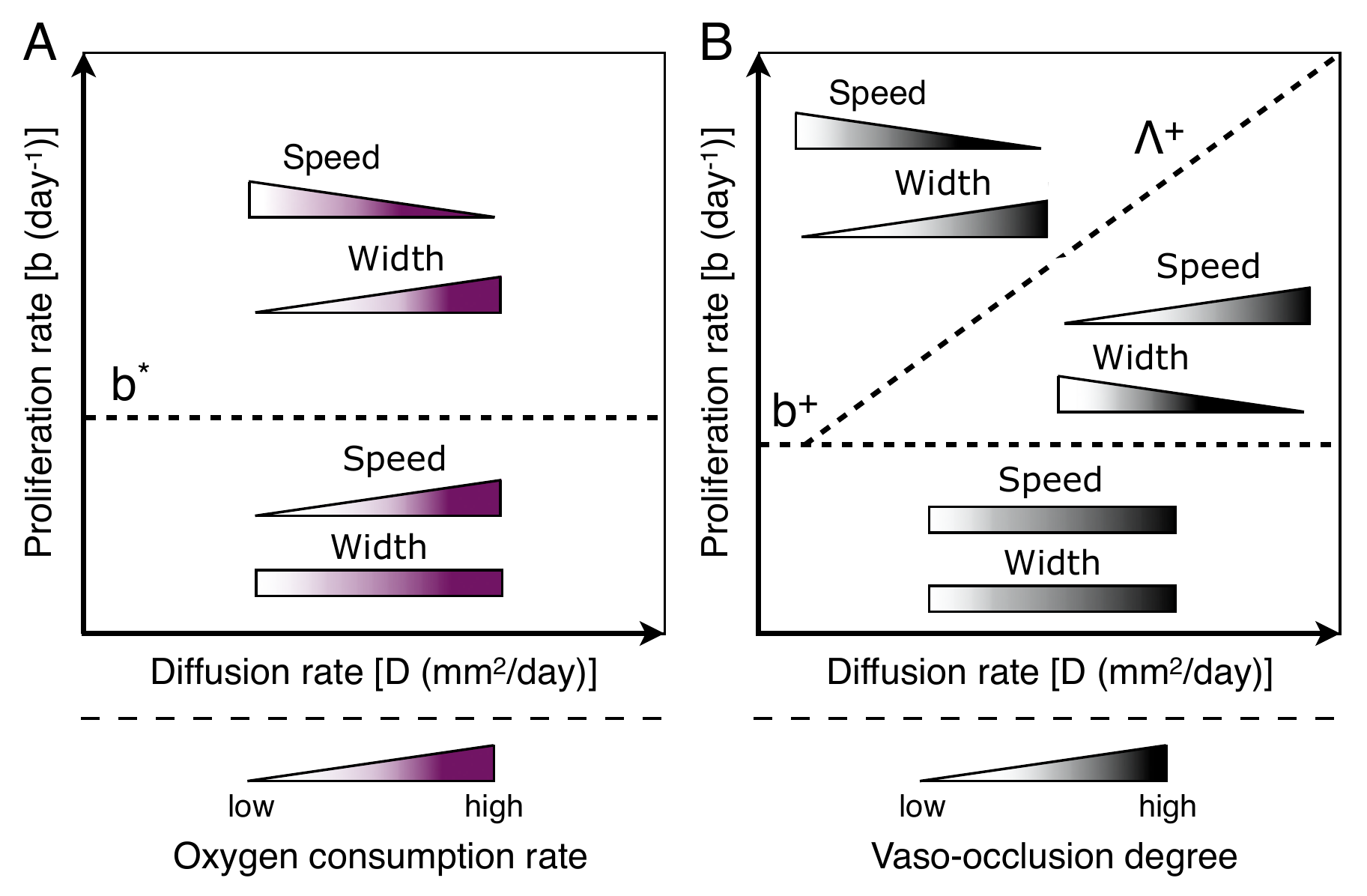}
\caption{\textbf{Overview of model simulation results.} (A) Modulations of cell oxygen consumption under the assumption of constant functional vasculature density reveal a critical proliferation rate $b^{*}$ in glioma invasion responses ({\it model II}). (B) Modulations of functional tumour-associated vasculature reveal a critical proliferation/diffusion ratio $\Lambda^{+} = b/D$ for proliferation rates higher than $b^{+}$ in glioma invasion responses ({\it model III}). Colour gradients from low to high represent the increase of cell oxygen consumption and vaso-occlusion rates. The purple and black wedges/bars represent the resulting effects on tumour front speed and infiltration width, for increasing/decreasing cell oxygen consumption and vaso-occlusion rates.}
\label{fig5}
\end{figure}

In this work, we developed a deterministic mathematical model of glioma invasion which is formulated as a system of reaction-diffusion equations. The model accounts for the dynamics of normoxic and hypoxic glioma cells based on the Go-or-Grow mechanism and influenced by the functional tumour-associated vasculature, as well as concentrations of pro-angiogenic factors and oxygen in the tumour microenvironment. Specifically, we focused on the effects of cell oxygen consumption and vascular modulations on relevant properties of glioma invasion, i.e. tumour front speed and infiltration width. The main simulation results of the model are summarised in Figure~\ref{fig5}.

The model analysis reveals that increasing cell oxygen consumption and vaso-occlusion rates result in more diffusive and less proliferative gliomas. In both scenarios, the average oxygen concentration in the tumour microenvironment decreases which limits cell proliferation and enhances hypoxia-induced migration. However, the extent to which such oxygen-mediated cell responses to vasculature-targeting treatment interventions influence glioma invasion depends on the specific intrinsic tumour features. Modulations of the functional tumour-associated vasculature reveals the existence of a critical cell proliferation/diffusion ratio for glioma invasion responses, see Figure~\ref{fig5}(B). This fact is observed for gliomas evolving with sufficiently high cell proliferation rates for variations in the oxygen concentration, due to vaso-occlusion or normalisation, significantly influences tumour cell dynamics. In such cases, tumour vascular modulations are predicted to produce opposing effects on front speed and infiltration width. Moreover, we found that depending on the intrinsic tumour features two distinct regimes can be identified where invasive behaviours in responses to vaso-modulatory interventions are different. A pro-thrombotic treatment is predicted to increase front speeds, but in turn reduces infiltration capacity of gliomas characterised by a cell proliferation/diffusion ratio below the critical threshold. On the contrary, gliomas in the other regime under the same treatment strategy become increasingly infiltrative and slowly growing. Analogously, vascular normalisation therapies produce opposing results for the corresponding parameter regimes.
\vspace{2mm}

Recently, it has been shown that the migration/proliferation dichotomy introduces a critical glioma cell density threshold separating tumour growth and extinction dynamics, a phenomenon called Allee effect \cite{Bottger2015}. Here, we also identify critical parameter values that distinguish different glioma invasive behaviours with respect to variations of cell oxygen consumption or vaso-occlusion. Interestingly, this is an emergent consequence of the Go-or-Grow plasticity, since in its absence (see \textit{model I}) no critical behaviour is observed. Assuming or not tumour vasculature dynamics, the Go-or-Grow induced criticality is expressed either in the form of a critical intrinsic proliferation/diffusion ratio $\Lambda^+$ or an intrinsic proliferation rate $b^*$, respectively, see Figure~\ref{fig5}. This result highlights the importance of further investigating the clinical effects of the Go-or-Grow phenomenon on glioma invasion.
\vspace{2mm}

The above \textit{in silico} findings demonstrate that {\it one-size-fits-all} vaso-modulatory interventions should be expected to fail to control glioma invasion due to the complexity of the involved mechanisms and the heterogeneity of patient- and tumour-related factors. This study proves the value of personalised treatment strategies based on a precise tumour profiling and provides a modelling framework with the potential to parametrise model predictions based on biopsy measurements. In particular, individual estimation of intrinsic proliferation and diffusion rates, for instance via biopsy tumour sample analysis, would be crucial components of such future tailored approaches to personalised glioma therapy. Moreover, this work substantially expands the current theoretical concepts in glioma invasion, showing that any vasculature-targeting therapeutic intervention will inevitably lead to a trade-off between tumour front speed and infiltration width. This finding suggests that vaso-modulatory therapies should be embedded in personalised combination therapy regimens, in which anti-angiogenesis might be integrated with individually adjusted other modules targeting proliferation, metabolism or tumour immunology. For instance, in the case of gliomas characterised by a high intrinsic proliferation/diffusion ratio, a pro-thrombotic or an anti-vasogenic treatment technique may reduce tumour invasion speed, but at the same time leads to highly infiltrative responses that makes this therapeutic strategy rather inappropriate. However, selecting a blood vessel normalisation strategy results in faster growing gliomas as a bulk with less-infiltrating morphologies. Thus, surgical resection could be considered to remove such compact tumours. In turn, the benefits of conventional treatments such as chemotherapy, radiotherapy and immunotherapy might increase in better-vascularised tumours \cite{Jain2001, Jain2005, Stylianopoulos2013, Jain2014a}. Therefore, an accurate glioma patient stratification during clinical decision-making is predicted relevant for the efficacy of vasculature-targeting therapies, based on either tumour-associated blood vessel deterioration or normalisation.
\vspace{2mm}

This work provides a mathematical framework for exploring novel approaches to rational combination therapies or regimens composed of subsequent periods of vaso-modulatory interventions and potentially other therapeutic modules. In our model the vaso-occlusion term is rather phenomenological and more accurate modelling is required. Furthermore, the migration/proliferation dichotomy has been modelled in the simplest possible way and more informed models could be integrated. In turn, intra-tumour genetic diversity is not directly considered, but we take into account phenotypic diversity depending on the oxygen availability, that is crucial for therapeutic outcomes. The latter is supported by evidences that genetic diversity is tumour-subtype specific and not significantly affected during treatment, while phenotypic heterogeneity is different before and after therapy \cite{Almendro2014}. Despite the fact that the model involves a large number of model parameters, their values were defined independently from each other based on published experimental data. For those parameters estimated, a parametric analysis was performed and we concluded that variations of their values do not affect the general conclusions of this study. At this stage, we restrict the modelling strategy to the effects of vasculature-targeting therapies, however, we are aware of the fact that further aspects of tumour biology may play a crucial role. In fact, we aim to investigate the interactions between the immune system and angiogenesis as an additional level of complexity given the potential success of immunomodulatory therapies. In particular, macrophages are likely to be involved in relevant mechanisms and will be included in future developments of the current approach. This is particularly relevant in the light of recent advanced in molecular classification of malignant gliomas \cite{Ceccarelli2016}. Mathematical modelling provides an integrative approach for conventional radiological, biopsy and molecular tumour characterization, allowing for the prediction of glioma treatment responses and translation into clinical decision-making. 

%--------------------------------------------------------------------------------------------------------------------------------------------------------------------------------------------------------------------------------------------------------------------%
%--------------------------------------------------------------------------------------------------------------------------------------------------------------------------------------------------------------------------------------------------------------------%
%--------------------------------------------------------------------------------------------------------------------------------------------------------------------------------------------------------------------------------------------------------------------%

\section*{Acknowledgments}
This work was partially supported by the Free State of Saxony and European Social Fund of the European Union (ESF, grant GlioMath-Dresden). J. C. L. Alfonso, F. Feuerhake and H. Hatzikirou gratefully acknowledge the funding support of the German Federal Ministry of Education and Research (BMBF) for the eMED project SYSIMIT (01ZX1308D). A. Deutsch acknowledges the support by Deutsche Krebshilfe. Authors also thank the Center for Information Services and High Performance Computing (ZIH) at TU Dresden for generous allocations of computational resources.

%--------------------------------------------------------------------------------------------------------------------------------------------------------------------------------------------------------------------------------------------------------------------%
%--------------------------------------------------------------------------------------------------------------------------------------------------------------------------------------------------------------------------------------------------------------------%
%--------------------------------------------------------------------------------------------------------------------------------------------------------------------------------------------------------------------------------------------------------------------%

\section*{Supplementary Material}
\vspace{2mm}

%--------------------------------------------------------------------------------------------------------------------------------------------------------------------------------------------------------------------------------------------------------------------%

\subsection*{1.1 Numerical implementation}

Numerical solutions of the proposed glioma-vasculature interplay model are obtained by implementing the finite element method and backward Euler scheme for spatial and temporal discretisation, respectively \cite{Larsson2008, Johnson2012}. The system of coupled partial differential equations (17)-(19) is first transformed into a weak formulation, which results in a system of ordinary differential equations with respect to time. The one-dimensional domain of simulation over which such equations are numerically solved is divided into a finite number of distinct and non-overlapping linear elements. The integrals involved in the weak form of the system are calculated on each domain element by means of a Gaussian quadrature formula, which exactly integrates the resulting polynomials \cite{Khursheed2012}. The backward Euler scheme is then used to obtain a temporal discretisation that results in a nonlinear system of equations solved at each instant of time by the Newton-Raphson method \cite{Larsson2008}. Model simulations were carried out using MATLAB software ({\it www.mathworks.com}) in a SuSE Linux Enterprise Server 11 with 5888 core AMD Opteron 6274 2.2GHz, 92 nodes each with 64 cores and 64 to 512 GB of memory.

%--------------------------------------------------------------------------------------------------------------------------------------------------------------------------------------------------------------------------------------------------------------------%

\subsection*{1.2 Simulation domain}

The system of equations (17)-(19) is solved in a one-dimensional domain $\Omega$ of length $L = 200$~mm for a total simulation time of 3 years, i.e. $T_{f} = 1095$~days. The independent system variables are time $t$ and space $x$ with $0 \leq x \leq L$ and $0 \leq t \leq T_f$. The $x$-axis can be thought of as a two-dimensional domain which is spatially averaged in one direction. The simulation domain, either inside the region occupied by glioma cells or outside representing the host brain tissue, is discretised into an irregular grid varying from a minimum segment length of $2.5 \times 10^{-3}$~mm to a maximum one of $2.5 \times 10^{-2}$~mm. The time step is taken equal to $0.25$~day, i.e. 6~hours. Both, segment length and time step are properly selected to ensure numerical stability.

%--------------------------------------------------------------------------------------------------------------------------------------------------------------------------------------------------------------------------------------------------------------------%

\subsection*{1.3 Model observables}

We characterise glioma invasion by the tumour front speed and infiltration width. The front speed is estimated by the rate of change given by the point of maximum slope in $\rho(x,t)$ at the end of numerical simulations $T_f$, see Figure~S1. In turn, the infiltration width is defined by the difference between the points where glioma cell density is $80\%$ and $2\%$ of the maximum cell density $\overline{\rho}$ at time $T_f$. 

\begin{figure}[H]
\centering \label{figure: diagrams}
\includegraphics[width=0.4\textwidth]{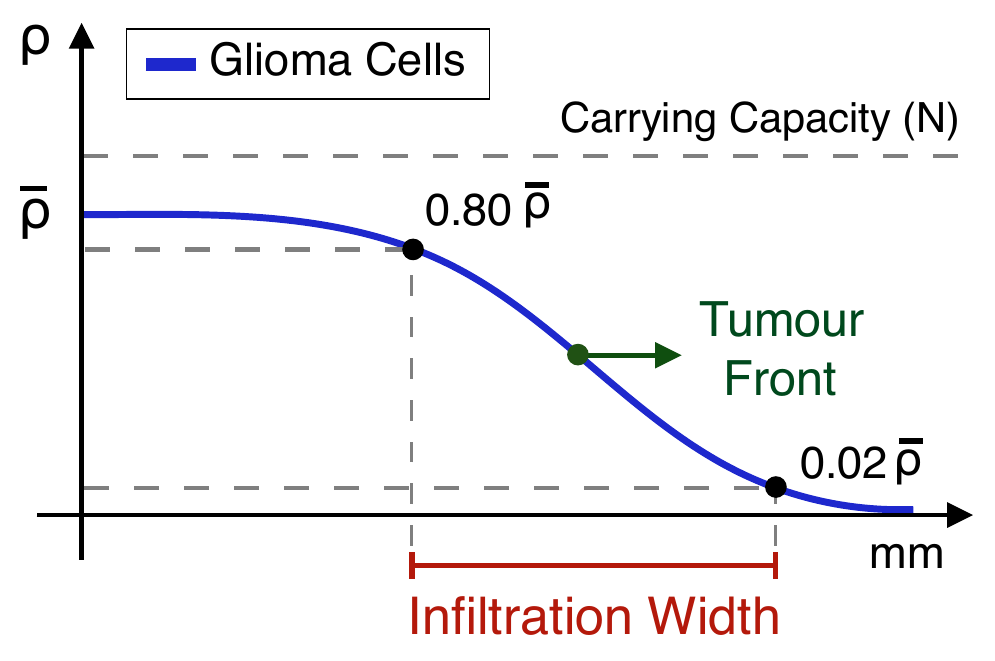}
\caption{Model observables. The tumour front is defined by the point of maximum slope in $\rho(x,t)$ (green) and infiltration width (red).}
\label{figS1}
\end{figure}

%--------------------------------------------------------------------------------------------------------------------------------------------------------------------------------------------------------------------------------------------------------------------%
%--------------------------------------------------------------------------------------------------------------------------------------------------------------------------------------------------------------------------------------------------------------------%
%--------------------------------------------------------------------------------------------------------------------------------------------------------------------------------------------------------------------------------------------------------------------%

\subsection*{2 Model parameterisation}

%--------------------------------------------------------------------------------------------------------------------------------------------------------------------------------------------------------------------------------------------------------------------%

\subsubsection*{2.1 Initial conditions}

Density of functional tumour vasculature and oxygen concentration are initialised in the domain $\Omega$ as $v_0 = 1/2$ and $\sigma_0 = 1.0$~nmol~mm$^{-1}$, respectively. In turn, the initial number of glioma cells, $p_0 = 40$~cells~mm$^{-1}$, is modulated by the continuous approximation of the Heaviside decreasing step function $\tilde{H}_{\gamma}(x-\epsilon) = 1 - \left(1/ \left( 1 + e^{-2 \gamma (x - \epsilon)} \right) \right)$ for $x \in \Omega$, with $\gamma = 1.0 \times 10^{1}$ and $\epsilon = 0.5$. The latter choice provides continuity on the model initial conditions and guarantees numerical stability. At both extremes of the simulation domain $\Omega$, no-flux boundary conditions are imposed.

%--------------------------------------------------------------------------------------------------------------------------------------------------------------------------------------------------------------------------------------------------------------------%
%--------------------------------------------------------------------------------------------------------------------------------------------------------------------------------------------------------------------------------------------------------------------%
%--------------------------------------------------------------------------------------------------------------------------------------------------------------------------------------------------------------------------------------------------------------------%

\subsubsection*{2.2 Density of glioma cells, $\rho(x, t)$}

\noindent \textbf{- Intrinsic diffusion rate of glioma cells $D$} (in mm$^2$~day$^{-1}$). Several studies using a data-driven Fisher-Kolmogorov model support that the diffusion rate of glioma cells is a patient-specific parameter \cite{Harpold2007, Swanson2008, Wang2009, Rockne2010, Swanson2011}. Estimates of $D$ vary from $2.73 \times 10^{-3}$ to $2.73 \times 10^{-1}$~mm$^2$~day$^{-1}$, which is supposed to cover low to high grade gliomas \cite{Harpold2007, Swanson2008, Swanson2011, Badoual2014}. Notice that $D_{\rho} = D/\alpha(\sigma_0)$, see equation (4) where $\alpha(\sigma)$ is defined.
\vspace{2mm}

\noindent \textbf{- Intrinsic proliferation rate of glioma cells $b$} (in day$^{-1}$). Similar as reported for the diffusion rate of glioma cells, $b$ is also suggested to be patient-specific \cite{Harpold2007, Swanson2008, Wang2009, Rockne2010, Swanson2011}. Estimates of $b$ vary from $2.73 \times 10^{-4}$ to $2.73 \times 10^{-2}$~day$^{-1}$, which is supposed to cover low to high grade gliomas \cite{Harpold2007, Swanson2008, Swanson2011, Badoual2014}. Notice that $b_{\rho} = b/\beta(\sigma_0)$, see equation (5) where $\beta(\sigma)$ is defined.
\vspace{2mm}

\noindent \textbf{- Brain tissue carrying capacity $N$} (in cells~mm$^{-1}$). This model parameter describes the limiting concentration of glioma cells that a volume of host brain tissue can hold. Considering an average glioma cell diameter of about 10~$\mu$m \cite{Swanson2011}, the one-dimensional carrying capacity is about $10^{2}$~cells~mm$^{-1}$. This estimate is in line with previous values considered for modelling of glioma growth \cite{Eikenberry2009, McDaniel2013}.
\vspace{2mm}

\noindent \textbf{- Physiological oxygen concentration in the host brain tissue $\sigma_0$} (in nmol~mm$^{-1}$). Although {\it in vivo} estimates of oxygen pressure in the brain tissue may vary with respect to measurement methods and other factors, a suitable experimental value for $\sigma_0$ is 40~mmHg \cite{Maas1993, Meixensberger1993, Hoffman1996, Carreau2011}. Henry's law \cite{Henry1803} is used to obtain the concentration of oxygen in the brain tissue as follows

\begin{equation*}
\sigma_0 = 40/k_H \approx 2.068~\mbox{nmol}~\mbox{mm}^{-3},
\end{equation*}

\noindent where $k_H = 1.93420922505 \times 10^{10}$~mm$^3$~mmHg~mol$^{-1}$ is the Henry's law constant for oxygen at normal body temperature.
\vspace{2mm}

\noindent To convert the three-dimensional oxygen concentration in the host brain tissue into its equivalent one-dimensional concentration, we multiply by the area of a transversal section of the tumour. We assume that such transversal section is equivalent to the surface area of a sphere of radius $r$, where $A = 4 \pi r^2$. Moreover, we consider that $r = 200~\mu\mbox{m} = 2.0 \times 10^{-1}$~mm is the characteristic nutrient diffusion length, which is consistent with the observed thickness of viable rims of tumour cells in spheroids \cite{Frieboes2006, Cristini2008, Hatzikirou2015}. Then, $A = 4 \pi (2 \times10^{-1})^2 = 16 \pi \times 10^{-2}$~mm$^2$ and we obtain that

\begin{equation*}
 \sigma_0 = 2.068~\mbox{nmol}~\mbox{mm}^{-3} \cdot (16 \pi \cdot 10^{-2}~\mbox{mm}^2)  \approx 1.0~\mbox{nmol}~\mbox{mm}^{-1}.
\end{equation*}

\noindent \textbf{- Phenotypic switching parameter (proliferative to migratory) $\lambda_1$} (in nmol~mm$^{-1}$). We take $\lambda_1 = \sigma_{M}$ in the phenotypic switching function $f_{21} = \lambda_1 - \sigma$ of glioma cells, where $\sigma_{M}$ is the maximum oxygen concentration in the host brain tissue. In normal brain tissues, oxygen tension has been estimated to range from 10 to 80~mmHg \cite{Luoto2013, Crawford2013}. Accordingly, we consider that $\sigma_{M} = 2.0$~nmol~mm$^{-1}$, i.e. for an oxygen pressure equal to 80~mmHg, which is two times higher than the assumed physiological oxygen concentration $\sigma_0$.
\vspace{2mm}

\noindent \textbf{- Phenotypic switching parameter (migratory to proliferative) $\lambda_2$} (dimensionless). The effect of $\lambda_2$ on glioma invasion is investigated by considering the following overall proliferation rate of glioma cells

\begin{equation*}
B = b \hspace{0.5mm} \frac{\beta(\sigma)}{\beta(\sigma_0)} = b \hspace{0.5mm} \frac{(\lambda_2 - 1)\sigma_0 + \lambda_1}{(\lambda_2 - 1)\sigma + \lambda_1} \frac{\sigma}{\sigma_0},
\end{equation*}

\noindent where taking into account that $\lambda_1 = \sigma_{M}$, we can distinguish the following three representative cases:
\vspace{2mm}

\hspace{10mm} (i) If $0 < \lambda_2 < 1$, then $B = b \hspace{0.5mm} \dfrac{\sigma_{M} - |\lambda_2 - 1| \sigma_0}{\sigma_{M} - |\lambda_2 - 1| \sigma} \hspace{0.5mm} \dfrac{\sigma}{\sigma_0} \propto \dfrac{\sigma}{\dfrac{\sigma_{M}}{|\lambda_2 - 1|} - \sigma}$.
\vspace{2mm}

\hspace{10mm} (ii) If $\lambda_2 = 1$, then $B = b \hspace{0.5mm} \dfrac{\sigma}{\sigma_0} \propto \sigma$.
\vspace{2mm}

\hspace{10mm} (iii) If $\lambda_2 > 1$, then $B = b \hspace{0.5mm} \dfrac{|\lambda_2 - 1| \sigma_0 + \sigma_{M}}{|\lambda_2 - 1| \sigma + \sigma_{M}} \hspace{0.5mm} \dfrac{\sigma}{\sigma_0} \propto \dfrac{\sigma}{\dfrac{\sigma_{M}}{|\lambda_2 - 1|} + \sigma}$.
\vspace{2mm}

According to (i)-(iii), we reduce model simulations to the following three parameter values $\lambda_2 = \{0.5,~1.0,~2.0\}$, see Figure~S2. Notice that in the limiting case of $\lambda_2 = 0$ glioma cells do not proliferate, and therefore we neglect this scenario. Although numerical simulations are obtained for the phenotypic switching parameter $\lambda_2 = 1.0$, we report in Section 3 the effect of $\lambda_2$ variations on glioma invasion, see Figures~S4 and S5.

\begin{figure}[H]
\centering \label{figure: diagrams}
\includegraphics[width=0.4\textwidth]{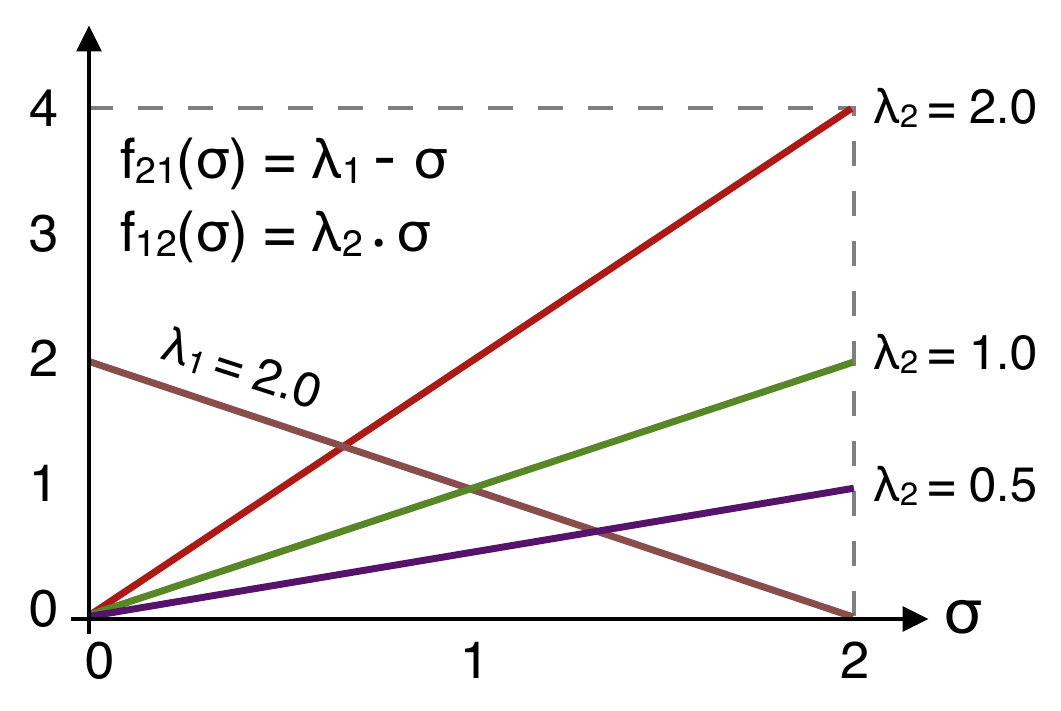}
\caption{Oxygen-dependant phenotypic switching functions based on the migration/proliferation dichotomy of glioma cells.}
\label{figS2}
\end{figure}

%--------------------------------------------------------------------------------------------------------------------------------------------------------------------------------------------------------------------------------------------------------------------%
%--------------------------------------------------------------------------------------------------------------------------------------------------------------------------------------------------------------------------------------------------------------------%
%--------------------------------------------------------------------------------------------------------------------------------------------------------------------------------------------------------------------------------------------------------------------%

\subsubsection*{2.3 Oxygen concentration, $\sigma(x, t)$}

\noindent \textbf{- Diffusion rate of oxygen $D_{\sigma}$} (in mm$^2$~day$^{-1}$). Based on experimental data, the oxygen diffusion rate in tumour tissues at 37~$^{\circ}$C has been reported equal to $1.75 \times 10^{-5}$~cm$^2$~s$^{-1}$ \cite{Grote1977}. Thus, we consider that $D_{\sigma} = 1.51 \times 10^{2}$~mm$^2$~day$^{-1}$, which is in agreement with previous estimates of the oxygen diffusion rate \cite{Matzavinos2009, Stamper2010, Powathil2012}.
\vspace{2mm}

\noindent \textbf{- Oxygen supply rate $h_1$} (in day$^{-1}$). Experimental estimates of transvascular permeability to oxygen $Pm_{O2}$ have been reported in the range $3 \times 10^{-5}$ to $3 \times 10^{-4}$~m~s$^{-1}$ \cite{Kelly2006}. In turn, the ratio of capillary surface area to volume $\frac{S}{V}$ has been observed to vary between 0.13 and 0.33~m$^{-1}$ \cite{Kelly2006}. Then, $Pm_{O2} \cdot \frac{S}{V}$ lies in the range $4.0 \times 10^{-6}$ to $1.0 \times 10^{-4}$~s$^{-1}$, which is equivalent to model parameter $h_1$ in the oxygen supply term of equation (15) \cite{Kelly2006}. These estimates are also in line with other oxygen supply rates reported \cite{Eggleton1998, Goldman2000}, i.e. $h_1 = 3.5 \times 10^{-6}$ and $4.0 \times 10^{-6}$~s$^{-1}$. Accordingly, we consider that $h_1 = 3.37 \times 10^{-1}$~day$^{-1}$, which is in the range of $h_1$ values above.
\vspace{2mm}

\noindent \textbf{- Oxygen consumption rate $h_2$} (in mm~cell$^{-1}$~day$^{-1}$). Oxygen consumption rates by tumour cells have been reported to vary from 2 to 40~$\mu$l~g$^{-1}$~min$^{-1}$ \cite{Vaupel1989, Grimes2014}. Considering the average mass of a cancer cell equal to $10^{-9}$~kg \cite{Powathil2012} and taking into account that $1~\mu\mbox{l} = 1$~mm$^3$, we have that $h_2$ is in the range $2.88 \times 10^{-3}$ to $5.76 \times 10^{-2}$~mm$^{3}$~cells$^{-1}$~day$^{-1}$. As explained above for the estimation of the physiological oxygen concentration in the host brain tissue $\sigma_0$, we convert the three-dimensional oxygen consumption rate into its equivalent one-dimensional rate dividing by the area of a transversal section of the tumour equal to $A = 16 \pi \cdot 10^{-2}$~mm$^2$. Thus, we have that $h_2$ is in the range $5.73 \times 10^{-3}$ to $1.14 \times 10^{-1}$~mm~cell$^{-1}$~day$^{-1}$.

%--------------------------------------------------------------------------------------------------------------------------------------------------------------------------------------------------------------------------------------------------------------------%
%--------------------------------------------------------------------------------------------------------------------------------------------------------------------------------------------------------------------------------------------------------------------%
%--------------------------------------------------------------------------------------------------------------------------------------------------------------------------------------------------------------------------------------------------------------------%

\subsubsection*{2.4 Density of functional tumour vasculature, $v(x, t)$}

\noindent \textbf{- Vasculature dispersal rate $D_{v}$} (in mm$^2$~day$^{-1}$). Experimental estimates of endothelial cell motility rate in different conditions have been reported between $10^{-3}$ and $10^{-4}$~mm$^2$~day$^{-1}$ \cite{Stokes1991, Kouvroukoglou2000}. We consider that $D_{v} = 5.0 \times 10^{-4}$~mm$^2$~day$^{-1}$, which is in line with previous models of vascularised tumour growth \cite{Anderson1998, Stamper2010, Swanson2011}.
\vspace{2mm}

\noindent \textbf{- Vasculature formation rate $g_1$} (in day$^{-1}$). We consider that $g_1 = 1.0 \times 10^{-1}$~day$^{-1}$ by assuming that blood vessels are formed in a timescale of hours \cite{Shaifer2010, Stamper2010, Scianna2013}. We remark that variations in the value of $g_1$ change the results only quantitatively, while qualitative phenomena are conserved. %We remark that parameter values of $g_1$ in the range $5.0 \times 10^{-2}$ to $5.0$~day$^{-1}$, i.e. timescale of minutes or days, change the results only quantitatively, while qualitative phenomena are conserved.
\vspace{2mm}

\noindent \textbf{- Oxygen concentration threshold for hypoxia $\sigma^{*}_{a}$} (in nmol~mm$^{-1}$). Although no consensus has been achieved for hypoxic thresholds, tumour tissues with $\mbox{P}_{\mbox{O2}}$ levels below 10~mmHg are usually considered under hypoxia \cite{Cardenas2004, Vaupel2007, Powathil2012}. Indeed, tissues with oxygen tension between 5.0 and 7.5~mmHg are considered under moderate hypoxia, and less than or equal to 2.5~mmHg under severe hypoxia \cite{Cardenas2004}. Accordingly, we assume that $\sigma^{*}_{a} = 2.5 \times 10^{-1}$~nmol~mm$^{-1}$, see also the derivation of the physiological oxygen concentration in the host brain tissue $\sigma_0$ for further details.
\vspace{2mm}

\noindent \textbf{- Half-maximal pro-angiogenic factor concentration $K$} (in nmol~mm$^{-1}$). We assume that the natural decay rate of pro-angiogenic factors is much smaller than the consumption rate by endothelial cells, i.e. $k_3 \ll k_2$ \cite{KohnLuque2013}. Then, taking into account the equation of effective pro-angiogenic factor concentration

\begin{equation*}
a = \frac{k_{1} \hspace{0.5mm} \rho \hspace{0.5mm} \tilde{\text{H}}_{\theta}(\sigma-\sigma_a^{*})}{k_{2}v+k_{3}},
\end{equation*}

\noindent the Michaelis-Menten kinetics on the density of functional tumour vasculature in equation (10) is as follows

\begin{equation*}
\frac{a}{\mu + a} = \dfrac{\dfrac{k_{1}\rho\tilde{\text{H}}_{\theta}(\sigma-\sigma^{*}_{a})}{k_{2}v+k_{3}}}{\mu+\dfrac{k_{1}\rho\tilde{\text{H}}_{\theta}(\sigma-\sigma^{*}_{a})}{k_{2}v+k_{3}}} =  \dfrac{ \dfrac{k_{1}}{k_{2}} \dfrac{\rho}{v}\tilde{\text{H}}_{\theta}(\sigma-\sigma^{*}_{a})} {\mu+\dfrac{k_{1}}{k_{2}} \dfrac{\rho}{v}\tilde{\text{H}}_{\theta}(\sigma-\sigma^{*}_{a})} = \dfrac{ \dfrac{\rho}{v}\tilde{\text{H}}_{\theta}(\sigma-\sigma^{*}_{a})} {K + \dfrac{\rho}{v}\tilde{\text{H}}_{\theta}(\sigma-\sigma^{*}_{a})},
\end{equation*}

\noindent where $K=\mu k_{2} / k_{1}$ is a positive constant denoting the concentration of pro-angiogenic factors at which the functional tumour vasculature formation rate is half-maximal. $\tilde{\text{H}}_{\theta}(\sigma-\sigma^{*}_{a})$ is a continuous approximation of the Heaviside decreasing step function $H(\xi)$, defined as $H(\xi) = 1$ if $\xi \leq 0$ and $H(\xi) = 0$ if $\xi > 0$, and given by

\begin{equation*}
\tilde{\text{H}}_{\theta}(\sigma-\sigma^{*}_{a}) = 1 - \frac{1}{1+e^{-2\theta (\sigma-\sigma^{*}_{a})}},
\end{equation*}

where $\theta = 1.0 \times 10^{1}$ and $K = 1.0 \times 10^{1}$~nmol~mm$^{-1}$. We remark that variations in the value of $K$ slightly change the results quantitatively, while qualitative phenomena are conserved. %We notice that values of $K$ in the range of $1.0$ to $1.0 \times 10^{2}$~nmol~mm$^{-1}$ change the results only quantitatively, while qualitative phenomena are conserved.
\vspace{2mm}

\noindent \textbf{- Vaso-occlusion term $G(v, \rho) = g_2 v \rho^n$}. Figure~S3(A) shows a schematic representation of vaso-occlusion, see also equation~(14). We assume that occlusion of tumour blood vessels only occurs for glioma cell densities greater than $N/2$, where $N$ is the brain tissue carrying capacity \cite{Padera2004}. Accordingly, we can distinguish the following two representative cases:
\vspace{2mm}

\hspace{10mm} (i) If $\rho \leq N/2$, \hspace{2mm} then \hspace{2mm} $G(v, \rho) = g_2 v \rho^n = g_2 \dfrac{N^n}{2^{n+1}} \approx 0$.\\

\hspace{10mm} (ii) If $\rho > N/2$, \hspace{1mm} then \hspace{2mm} $G(v, \rho) = g_2 v \rho^n > g_2 \dfrac{N^n}{2^{n+1}} > 0$.\\

Considering the functional tumour vasculature at normal density, i.e. $v = 1/2$, we have that to satisfy the above assumption on vaso-occlusion induced by glioma cell density low and high values of $g_2$ and $n$ are required, respectively. Therefore, we take $n = 6$ and consider the following values of $g_2 = \{5.0 \times 10^{-13},~5.0 \times 10^{-12},~1.5 \times 10^{-11}\}$~cell$^{-n}$~mm$^{n}$~day$^{-1}$. We remark that lower values of $n$ do not reproduce the experimental observation that vaso-occlusion starts to occur at tumour cell densities greater than $N/2$. Figure~S3(B) shows the dependence of the vaso-occlusion term $G(v, \rho)$ in equation~(14) on the density of glioma cells $\rho$ for $n = 6$, $v = 1/2$ and values of $g_2$ considered. In turn, Figure~S3(C) provides simulation maps of the vaso-occlusion percentage depending on the diffusion and proliferation rates of glioma cells at the end of numerical simulations $T_{f} = 3$~years. This percentage is obtained as the ratio between the integral of $v(x,T_f)$ from $x = 0$ to the point $x_v$ where $v = 1/2$ and the area of the rectangle given by $x_v/2$. We observe that vaso-occlusion increases as the proliferation rate of glioma cells becomes higher, see Figure~S3(C).
\vspace{2mm}

The term $G(v, \rho) = g_2 v \rho^n$ in equation~(14) is selected to model vaso-occlusion because from our experience extensive tumour blood vessel collapse is taking place when solid stress exceeds a critical value \cite{Stylianopoulos2013a, Stylianopoulos2013b}. Prior to this critical stress threshold, blood vessel collapse is moderate \cite{Stylianopoulos2013b}. We remark that the use of a different expression for $G(v, \rho)$ would change the results only quantitatively and it is not expected to affect the general conclusions of this study.

\begin{figure}[H]
\centering
\includegraphics[width=0.7\textwidth]{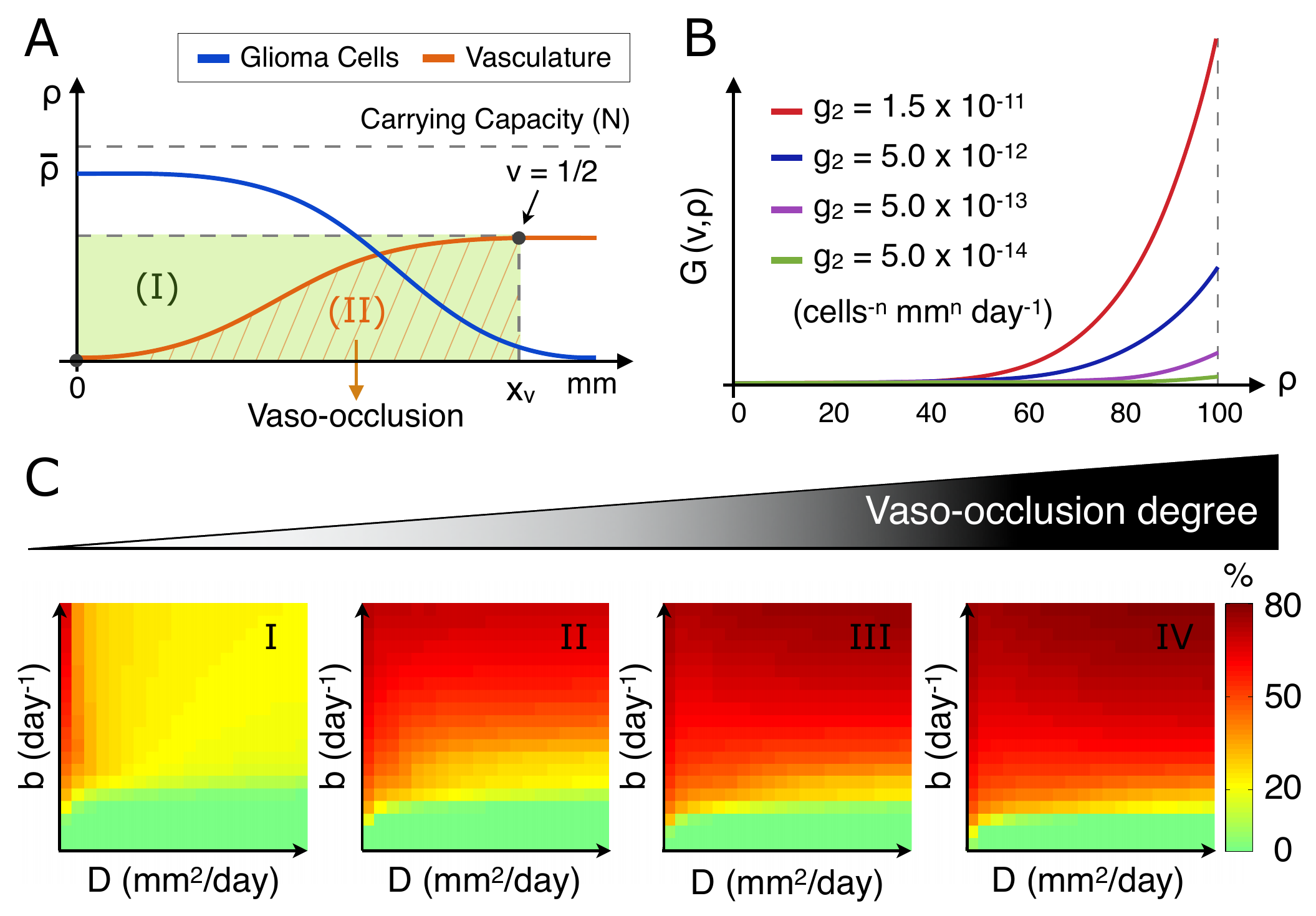}
\caption{(A) Schematic representation of vaso-occlusion. (B) Dependence of the vaso-occlusion term $G(v, \rho) = g_2 v \rho^n$ on density of glioma cells $\rho$ for $v = 1/2$, $n = 6$ and different values of $g_2$. (C) Vaso-occlusion percentage, at the end of numerical simulations $T_{f} = 3$~years, with respect to intrinsic diffusion and proliferation rates of glioma cells for a fixed oxygen consumption rate $h_2 = 5.73 \times 10^{-3}$~mm~cell$^{-1}$~day$^{-1}$ and $g_2 = \{5.0 \times 10^{-14},~5.0 \times 10^{-13},~5.0 \times 10^{-12},~1.5 \times 10^{-11}\}$~cells$^{-n}$~mm$^{n}$~day$^{-1}$ in simulation maps I-IV, respectively. Other model parameters are as in Table~1.}
\label{figS3}
\end{figure}

%--------------------------------------------------------------------------------------------------------------------------------------------------------------------------------------------------------------------------------------------------------------------%
%--------------------------------------------------------------------------------------------------------------------------------------------------------------------------------------------------------------------------------------------------------------------%
%--------------------------------------------------------------------------------------------------------------------------------------------------------------------------------------------------------------------------------------------------------------------%

\subsection*{3 Effect of phenotypic switching parameter $\lambda_2$ on model observables}

Numerical simulations are obtained for the phenotypic switching parameter $\lambda_2 = 1.0$. Thus, in order to complete the model analysis we investigate the effect of different values of $\lambda_2 = \{0.5,~1.0,~2.0\}$ on glioma invasion. Indeed, this set of $\lambda_2$ values covers the three representative cases discussed above, see also Figure~S2. As shown in Figures~S4 and S5, for increasing values of $\lambda_2$ the tumour front speed increases, while the infiltration width decreases. We further note that such changes in glioma invasion are similar with respect to the intrinsic tumour features. Based on these results, we can state that glioma invasion in response to variations of $\lambda_2$ is only quantitatively influenced, while qualitative phenomena are conserved.

\begin{figure}[H]
\centering 
\includegraphics[width=0.9\textwidth]{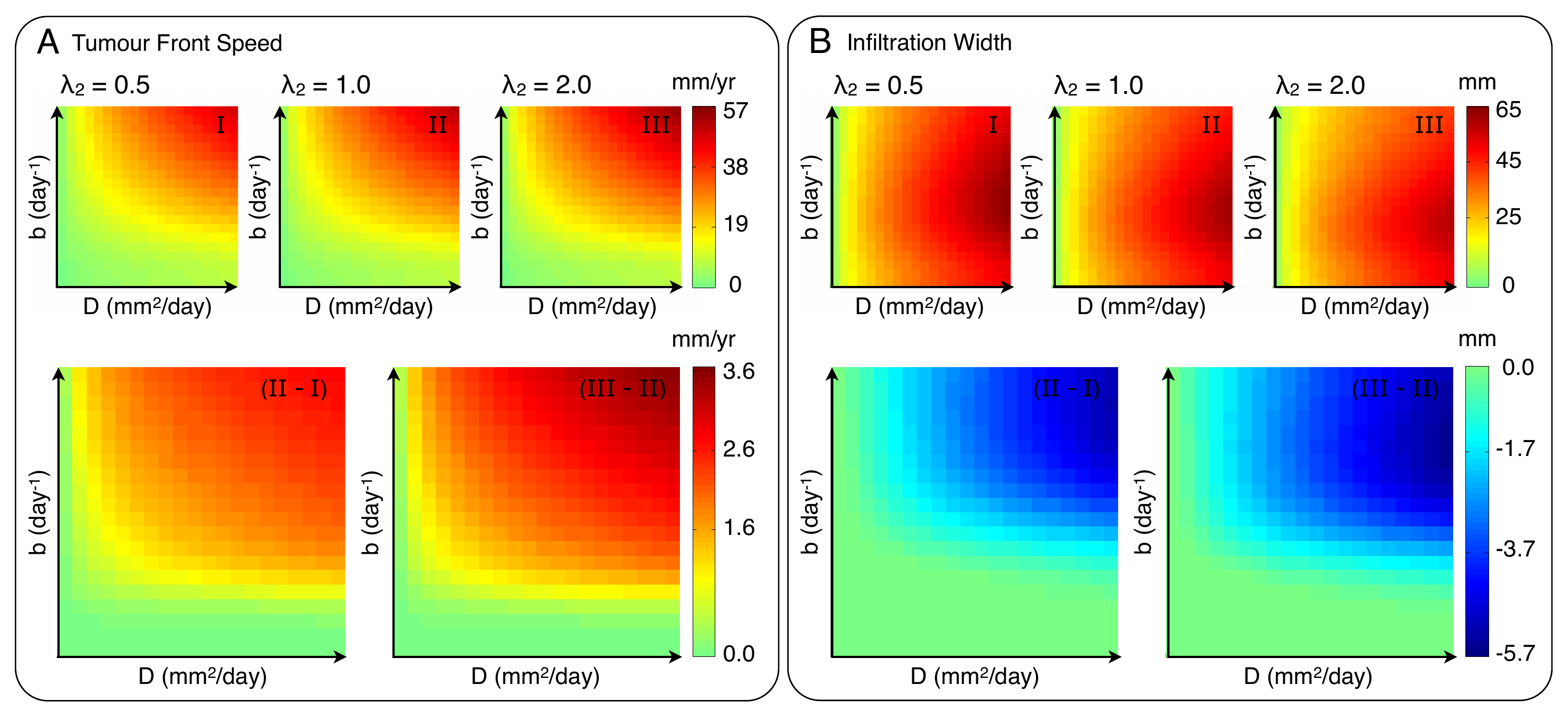}
\caption{\textbf{Model observables with respect to parameter $\lambda_2$ for constant functional tumour vasculature.} Simulation maps with respect to the intrinsic proliferation $b \in [2.73 \times 10^{-4},~2.73 \times 10^{-2}]$~days$^{-1}$ and diffusion $D \in [2.73 \times 10^{-3},~2.73 \times 10^{-1}]$~mm$^2$~days$^{-1}$ rates of glioma cells. (A) tumour front speed and (B) infiltration width for a fixed oxygen consumption $h_2 = 5.73 \times 10^{-3}$~mm~cell$^{-1}$~day$^{-1}$ rate, and values of $\lambda_2 = \{0.5,~1.0,~2.0\}$ in simulation maps I-III, respectively. (A-B) Differences between simulation maps are provided. The other parameters are as in Table~1.}
\label{figS4}
\end{figure}

\begin{figure}[H]
\centering 
\includegraphics[width=0.9\textwidth]{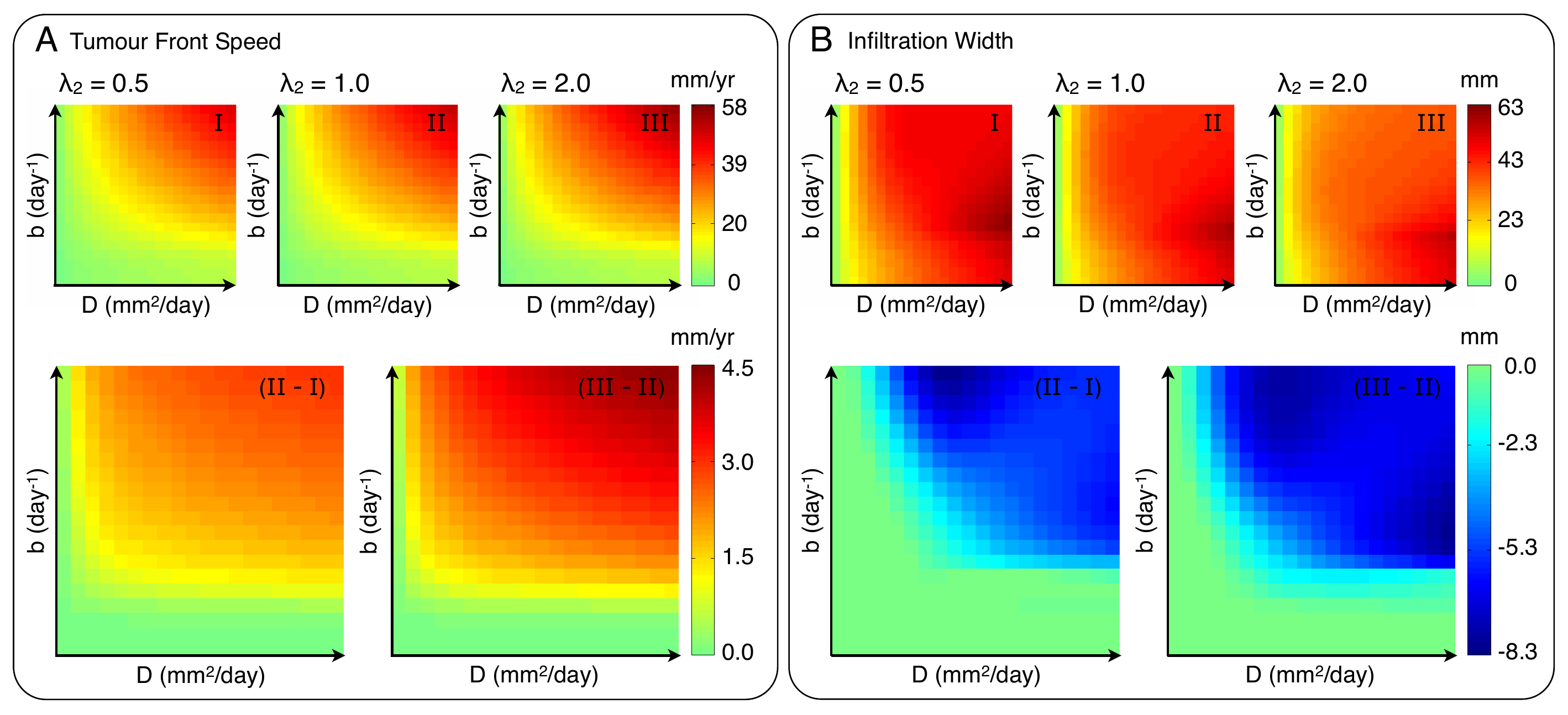}
\caption{\textbf{Model observables with respect to parameter $\lambda_2$.} Simulation maps with respect to the intrinsic proliferation $b \in [2.73 \times 10^{-4},~2.73 \times 10^{-2}]$~days$^{-1}$ and diffusion $D \in [2.73 \times 10^{-3},~2.73 \times 10^{-1}]$ mm$^2$ days$^{-1}$ rates of glioma cells. (A) tumour front speed and (B) infiltration width for fixed oxygen consumption $h_2 = 5.73 \times 10^{-3}$~mm~cell$^{-1}$~day$^{-1}$ and vaso-occlusion $g_2 = 5.0 \times 10^{-12}$~cells$^{-n}$~mm$^{n}$~day$^{-1}$ rates, and values of model parameter $\lambda_2 = \{0.5,~1.0,~2.0\}$ in simulation maps I-III, respectively. (A-B) Differences between simulation maps are provided. The other parameters are as in Table~1.}
\label{figS5}
\end{figure}

%--------------------------------------------------------------------------------------------------------------------------------------------------------------------------------------------------------------------------------------------------------------------%
%--------------------------------------------------------------------------------------------------------------------------------------------------------------------------------------------------------------------------------------------------------------------%
%--------------------------------------------------------------------------------------------------------------------------------------------------------------------------------------------------------------------------------------------------------------------%
 
{\footnotesize
\bibliographystyle{ieeetr}
%\bibliography{gliovasc_draft}}

}

\end{document}